# NEW EXACT SOLUTIONS OF THE 3D SCHRÖDINGER EQUATION

**E.E. Perepelkin[a,c*], B.I. Sadovnikov[a], N.G. Inozemtseva[b], A.S. Medvedev[a]**

[a] *Faculty of Physics, Lomonosov Moscow State University, Moscow, 119991 Russia*
[b] *Moscow Technical University of Communications and Informatics, Moscow, 123423 Russia*
[c] *Joint Institute for Nuclear Research, Moscow region, Dubna,141980 Russia*
**Corresponding author: pevgeny@jinr.ru*

**Abstract**

Previously we found a unique quantum system with a positive gauge-invariant Weyl-Stratonovich quasi-probability density function which can be defined by the so-called «quadratic funnel» potential. In this work we have constructed a class of exact solutions to the 3D Schrödinger equation for a two-parameter «quadratic funnel» potential based on the $\Psi$-model of micro and macro systems. It should be noted that the Aharonov-Bohm effect is inherent in the quantum $\Psi$-model. Explicit expressions for the energy spectrum and the set of eigenfunctions have been found. Using gauge invariance for scalar and vector potentials, a solution to the electromagnetic Schrödinger equation has been obtained, with a magnetic field in the form of a «Dirac string» defined by a singular vortex probability flux field.

Superpositions of eigenfunctions leading to various types of vortex and potential probability current fields have been investigated in detail. The analysis of the quantum system's properties has been carried out within the Wigner-Vlasov formalism.

## Introduction

Quantum mechanics has always remained the most non-trivial and unusual branch of physics. There exist at least nine different formulations of it [1]. Research has shown that the behavior of quantum objects is sometimes difficult to fit within the framework of macro-world logic. Today, with the development of technologies and the emergence of fields such as quantum computing [2-5], quantum information science [6-9], quantum communication [10-12], synthesis and practical use of quantum dots [13-15], quantum mechanics has become significantly «closer» in practical applied research.

The design of any complex modern system is associated with mathematical modeling. The availability of exact analytical solutions to fundamental equations has several advantages. First, a broad methodological aspect that allows studying the basic principles of operation, structure, and construction of such systems. Second, when creating software packages that simulate the behavior of complex nonlinear systems, it is important to have exact analytical solutions for testing algorithms. It is desirable to have as many such examples as possible to reduce the likelihood of numerical algorithm malfunction.

The Schrödinger equation is a fundamental equation of quantum mechanics, for which not many exact solutions are known [16-23]. The goal of this work was to construct an exact 3D solution to the Schrödinger equation for a two-parameter $(p_1, p_2)$ potential

$$U^{(p_1,p_2)}(r,\theta) = \frac{\hbar^2}{8m\sigma_r^2}\left[\frac{r^2}{\sigma_r^2} - p_1\frac{\sigma_r^2}{r^2} - p_2\frac{\sigma_r^2}{r^2\sin^2\theta}\right], \tag{i.1}$$

where $(r,\theta,\phi)$ is the spherical coordinate system, $\hbar$ is the reduced Planck constant, $m$ is the mass, and $\sigma_r$ is a constant value corresponding to the characteristic size of the system.

Potential (i.1) is an extended version of the potential $U^{(0,p_2)}$ considered in [24] when studying the properties of the gauge-invariant Weyl-Stratonovich function defining the quasi-probability density of a quantum system in phase space [25-29]. The unconventional consideration of quantum systems in phase space was initiated in the works of Weyl and Wigner [30, 31] back in 1932 by introducing the quasi-probability density function

$$W(\vec{r},\vec{p},t)=\frac{1}{(2\pi\hbar)^3}\int_{\mathbb{R}^3}\left\langle \vec{r}+\frac{\vec{s}}{2}\middle|\hat{\rho}\middle|\vec{r}-\frac{\vec{s}}{2}\right\rangle e^{-\frac{i}{\hbar}\vec{s}\cdot\vec{p}}\,d^3s, \tag{i.2}$$

where $\hat{\rho}$ is the density matrix. The «quantum» peculiarity of the Wigner function is the presence of regions in phase space (of size $\sim\hbar$) where the probability density takes negative values. Despite its specific properties, function (i.2) has found wide application in applied problems [32-35], allowing the calculation of phase space averages of various characteristics of a quantum system.

In Hudson's work [36], a theorem was proved for the 1D case that the Wigner function is positive throughout the entire phase space only for the Gaussian wave function. Later, in [37], this statement was extended to 3D systems. The evolution of function (i.2) is described by the Moyal equation (i.3), which is a quantum analogue of the Liouville equation from classical mechanics

$$\frac{\partial W}{\partial t}+\frac{1}{m}\vec{p}\cdot\nabla_r W-\nabla_r U\cdot\nabla_p W=\sum_{l=1}^{+\infty}\frac{(-1)^l(\hbar/2)^{2l}}{(2l+1)!}U\left(\overleftarrow{\nabla}_r\cdot\overrightarrow{\nabla}_p\right)^{2l+1}W, \tag{i.3}$$

where the arrow above the operator indicates the direction of its action.

When considering systems with electromagnetic interaction described by the Schrödinger equation

$$i\hbar\frac{\partial\Psi}{\partial t}=\frac{1}{2m}\left(\hat{p}-q\vec{A}\right)^2\Psi+U\Psi, \tag{i.4}$$

it was found that the Wigner function (i.2) is not gauge-invariant under the transformation [38]

$$\tilde{U}=U-\hbar\frac{\partial\tilde{\varphi}}{\partial t},\qquad -\hbar\nabla_r\tilde{\varphi}=q\vec{A},\qquad \Psi=\tilde{\Psi}e^{-i\tilde{\varphi}}, \tag{i.5}$$

where $\hat{p}=-i\hbar\nabla_r$ is the momentum operator, $\vec{A}$ is the vector potential of the magnetic field $\vec{B}=\operatorname{curl}_r\vec{A}$. As a result, a new form of quasi-density function [27-29] was proposed, possessing gauge invariance

$$f_w(\vec{r},\vec{P},t)=\frac{1}{(2\pi\hbar)^3}\int_{\mathbb{R}^3}e^{-\frac{i}{\hbar}\vec{s}\cdot\left[\vec{P}+\frac{q}{2}\int_{-1}^{1}\vec{A}\left(\vec{r}+\tau\frac{\vec{s}}{2},t\right)d\tau\right]}\rho\left(\vec{r}+\frac{\vec{s}}{2},\vec{r}-\frac{\vec{s}}{2},t\right)d^3s, \tag{i.6}$$

where $\vec{P}=\vec{p}-q\vec{A}$ and $\rho(\vec{r}_+,\vec{r}_-,t)=\Psi^*(\vec{r}_-,t)\Psi(\vec{r}_+,t)$, $\vec{r}_\pm=\vec{r}\pm\vec{s}/2$.

In [24], it was shown that function (i.6) is not positive for the Gaussian wave functions. Moreover, a wave function $\Psi_f$ different from the Gaussian distribution was found that gives positivity to the Weyl-Stratonovich function (i.6). It turned out that such a wave function $\Psi_f$ is a particular solution of the so-called $\Psi$- model constructed in [39]. Thus, the $\Psi$- model with potential $U^{(0,p_2)}$ led to a unique wave function $\Psi_f$ giving positivity to the quasi-probability density (i.6). In [24], the function $\Psi_f$ corresponded to the ground state of a quantum system with potential $U^{(0,p_2)}$. It seems natural to desire to find the complete energy spectrum and eigenfunctions of such a system. In this work, within the $\Psi$- model, this problem was solved for an extended two-parameter potential $U^{(p_1,p_2)}$.

The paper has the following structure. Section 1 describes the Wigner-Vlasov formalism, which provides a connection between classical and quantum systems through the chain of Vlasov equations [40, 41]. Based on this formalism, the construction of the $\Psi$- model [39] and its possible particular solutions are described. Sections 2-3 present the finding of exact singular and bounded solutions to the 3D Schrödinger equation with potential (i.1). Superpositions of eigenfunctions giving various types of vortex and potential probability current fields leading to the system's intrinsic magnetic moment are also considered. The conclusion briefly formulates the main results of the work. The Appendix contains proofs of theorems and intermediate mathematical transformations.

## §1 Wigner-Vlasov formalism

Within the framework of this article, we consider only the first two equations from the infinite self-consistent chain of Vlasov equations [40-42]

$$\hat{\pi}_n S_n = -Q_n,\ n = 1,2, \tag{1.1}$$

$$\hat{\pi}_1 \overset{\text{def}}{=} \frac{\partial}{\partial t} + \langle \vec{v} \rangle \cdot \nabla_r,\ \hat{\pi}_2 \overset{\text{def}}{=} \frac{\partial}{\partial t} + \vec{v} \cdot \nabla_r + \langle \dot{\vec{v}} \rangle \cdot \nabla_v, \tag{1.2}$$

$$S_n \overset{\text{def}}{=} \operatorname{Ln} f_n,\quad Q_1 \overset{\text{def}}{=} \operatorname{div}_r \langle \vec{v} \rangle,\ Q_2 \overset{\text{def}}{=} \operatorname{div}_v \langle \dot{\vec{v}} \rangle, \tag{1.3}$$

where

$$1 = \int_{\mathbb{R}^3} f_1(\vec{r},t)\, d^3r = \int_{\mathbb{R}^3}\int_{\mathbb{R}^3} f_2(\vec{r},\vec{v},t)\, d^3r d^3v, \tag{1.4}$$

$$f_1(\vec{r},t)\langle \vec{v} \rangle(\vec{r},t) = \int_{\mathbb{R}^3} f_2(\vec{r},\vec{v},t)\vec{v} d^3v,\quad f_1(\vec{r},t)\langle\langle \dot{\vec{v}} \rangle\rangle(\vec{r},t) = \int_{\mathbb{R}^3} f_2 \langle \dot{\vec{v}} \rangle d^3v, \tag{1.5}$$

$$f_2(\vec{r},\vec{v},t)\langle \dot{\vec{v}} \rangle(\vec{r},\vec{v},t) = \int_{\mathbb{R}^3} f_3(\vec{r},\vec{v},\dot{\vec{v}},t)\dot{\vec{v}} d^3\dot{v}.$$

The functions $f_1$ and $f_2$ are probability densities for coordinate and phase space, respectively, and satisfy the normalization condition (1.4). The kinematic quantities (1.5) define the average velocity vector fields and acceleration of the probability flux for coordinate and phase space. The differential operators (1.2) are essentially total time derivatives along the phase trajectory/characteristic. The quantities $Q_n$ (1.3) in equation (1.1) specify the density of dissipation sources. Thus, from equations (1.1) it follows that the total time derivative along the trajectory equals the density of dissipation sources with the opposite sign. If there are no dissipation sources $Q_n = 0$, then the probability density is conserved along the trajectory. For example, when $Q_2 = 0$, the second Vlasov equation $\hat{\pi}_2 S_2 = 0$ transforms into the well-known

Liouville equation for function $f_2$, where the first Vlasov approximation $m\langle\dot{\vec{v}}\rangle = -\nabla_r U$ [43] is used.

The equations (1.1) are related to the $H_n$ - functions of Boltzmann [44, 42]

$$H_1(t) \overset{\text{def}}{=} \int_{\mathbb{R}^3} f_1(\vec{r},t) S_1(\vec{r},t)\, d^3r = \langle S_1\rangle(t),\ H_2(t) \overset{\text{def}}{=} \int_{\mathbb{R}^3}\int_{\mathbb{R}^3} f_2(\vec{r},\vec{v},t) S_2(\vec{r},\vec{v},t)\, d^3r d^3v = \langle\langle S_2\rangle\rangle(t),$$

$$\hat{\pi}_0[H_1] = -\langle Q_1\rangle,\ \hat{\pi}_0[H_2] = -\langle\langle Q_2\rangle\rangle, \tag{1.6}$$

where $\hat{\pi}_0 \overset{\text{def}}{=} d/dt$. From equations (1.6) it follows that the average dissipation sources determine the rate of change of the Boltzmann $H_n$ - functions.

The second Vlasov equation ($n=2$) corresponds to the conservation laws of probability, momentum, and energy

$$\hat{\pi}_1\langle v_k\rangle = \left(\frac{\partial}{\partial t} + \langle v_\lambda\rangle \frac{\partial}{\partial x^\lambda}\right)\langle v_k\rangle = -\frac{1}{f_1}\frac{\partial P_{k\lambda}}{\partial x^\lambda} + \langle\langle \dot{v}_k\rangle\rangle, \tag{1.7}$$

$$\begin{aligned}&\frac{\partial}{\partial t}\left[\frac{f_1}{2}|\langle\vec{v}\rangle|^2 + \frac{1}{2}\operatorname{Tr} P_{kk}\right] + \\ &+\frac{\partial}{\partial x^\lambda}\left[\frac{f_1}{2}|\langle\vec{v}\rangle|^2\langle v_\lambda\rangle + \frac{1}{2}\langle v_\lambda\rangle \operatorname{Tr} P_{kk} + \langle v_k\rangle P_{k\lambda} + \frac{1}{2}\operatorname{Tr} P_{kk\lambda}\right] = \int_{\mathbb{R}^3} f_2\langle\dot{v}_k\rangle v_k\, d^3v,\end{aligned} \tag{1.8}$$

$$P_{k\lambda} = \int_{\mathbb{R}^3} f_2\left(v_k - \langle v_k\rangle\right)\left(v_\lambda - \langle v_\lambda\rangle\right) d^3v,\ P_{k\lambda s} \overset{\text{det}}{=} \int_{\mathbb{R}^3} f_2\left(v_k - \langle v_k\rangle\right)\left(v_\lambda - \langle v_\lambda\rangle\right)\left(v_s - \langle v_s\rangle\right) d^3v. \tag{1.9}$$

The law of conservation of probability/mass/charge is essentially the first Vlasov equation, which is obtained from the second by integrating over velocity space. The laws of conservation of momentum (1.7) and energy (1.8) are obtained by multiplying the second equation by the velocity component $v_\lambda$ and $v^2/2$, respectively, followed by integration over velocity space [41, 42]. On the left side of the equation of motion (1.7) is the derivative of velocity, and on the right is essentially the force due to external influence $\langle\langle\dot{v}_k\rangle\rangle$ and pressure force $\sim \partial P_{k\lambda}/\partial x^\lambda$. The first term on the left side of equation (1.8) determines the change in energy density over time, and the second term corresponds to the divergence of the energy flux density. On the right side of equation (1.8) is the work of external forces acting on the system.

Note that the Vlasov equation chain is self-consistent, so to find a solution, it is necessary to cut it off at some equation by introducing an approximation of the average kinematic quantity. Let us consider cutting the chain off at the first equation. According to the Helmholtz theorem, the vector field of the probability flux $\langle\vec{v}\rangle$ can be represented as a sum of potential $\nabla_r\Phi$ and vortex fields $\vec{A}$

$$\langle\vec{v}\rangle(\vec{r},t) = -\alpha\nabla_r\Phi(\vec{r},t) + \gamma\vec{A}(\vec{r},t), \tag{1.10}$$

where $\alpha,\gamma$ are some constant values. In [45], using expansion (1.10) and the representation of probability density $f_1 = |\Psi|^2$, $\Psi = |\Psi|\exp(i\varphi)$, $\Phi = 2\varphi + 2\pi k, k \in \mathbb{Z}$, the equation was obtained

$$\frac{i}{\beta}\frac{\partial \Psi}{\partial t} = -\alpha\beta\left(\hat{\mathrm{p}} - \frac{\gamma}{2\alpha\beta}\vec{A}\right)^2 \Psi + U\Psi, \tag{1.11}$$

$i\beta\hat{\mathrm{p}} \overset{\mathrm{def}}{=} \nabla_r$, which transforms into the electromagnetic Schrödinger equation when $\alpha = -\hbar/2m$, $\beta = 1/\hbar$, $\gamma = -q/m$. Substituting the representation $\Psi = |\Psi|\exp(i\varphi)$ into equation (1.11) leads to the Hamilton-Jacobi equation

$$-\frac{1}{\beta}\frac{\partial \varphi}{\partial t} = -\frac{1}{4\alpha\beta}\left|\langle \vec{v}\rangle\right|^2 + \mathrm{V} \overset{\mathrm{def}}{=} \mathrm{H}, \qquad \mathrm{V} = U + \mathrm{Q}, \quad \mathrm{Q} = \frac{\alpha}{\beta}\frac{\Delta_r|\Psi|}{|\Psi|}, \tag{1.12}$$

where $\mathrm{Q}$ is the quantum potential [46-47]. Applying the operator $\nabla_r$ to equation (1.12) and taking into account expansion (1.10) gives the equation of motion of a charged particle in an electromagnetic field

$$\hat{\pi}_1\langle \vec{v}\rangle = \frac{d}{dt}\langle \vec{v}\rangle = -\gamma\left(\vec{E} + \langle \vec{v}\rangle \times \vec{B}\right), \tag{1.13}$$

$$\vec{E} \overset{\mathrm{def}}{=} -\frac{\partial \vec{A}}{\partial t} - \frac{2\alpha\beta}{\gamma}\nabla_r \mathrm{V}, \qquad \vec{B} \overset{\mathrm{def}}{=} \frac{1}{\gamma}\mathrm{curl}_r\langle \vec{v}\rangle = \mathrm{curl}_r \vec{A}. \tag{1.14}$$

Cutting the Vlasov chain off at the second equation can be done by introducing the Vlasov-Moyal approximation for the average acceleration flux [43]

$$f_2\langle \dot{v}_k\rangle = \sum_{l=0}^{+\infty}\frac{(-1)^{l+1}(\hbar/2)^{2l}}{m^{2l+1}(2l+1)!}\frac{\partial U}{\partial x^k}\left(\overleftarrow{\nabla}_r \cdot \vec{\nabla}_v\right)^{2l} f_2. \tag{1.15}$$

Substituting approximation (1.15) into the second ($n = 2$) Vlasov equation (1.1) transforms it into the Moyal equation (i.3) for the Wigner function (i.2). Note that when deriving the Vlasov chain, no positivity condition was imposed on the functions $f_n$, so the negative values of the Wigner function do not contradict its basis. Approximation (1.15) can be directly used in the conservation law (1.8). To substitute (1.15) into the equation of motion (1.7), it is necessary to average it over velocities (1.5). As a result, the first Vlasov approximation $m\langle\langle \dot{\vec{v}}\rangle\rangle = -\nabla_r U$, introduced by him empirically, will be obtained. Thus, quantum corrections (terms with $\hbar^{2l}$) in series (1.15) vanish when averaging over velocities.

The above expressions allow us to connect classical and quantum systems within a single mathematical formalism of Wigner-Vlasov. Knowing the solutions to the Vlasov equations, we can construct solutions to the Schrödinger equation. As an example, let us consider an exact solution for the $\Psi$ - model of micro and macro systems. We take the expression for the scalar potential in the form

$$\varphi(\theta,\phi,t) = \ell\phi + \tau\theta - \frac{\mathrm{E}}{\hbar}t, \tag{1.16}$$

where $\ell, \tau, \mathrm{E}$ are constant values. According to (1.10), the potential (1.16) corresponds to the probability flux velocity field

$$\langle\vec{v}\rangle^{(\ell,\tau)}(r,\theta)=\frac{\hbar}{m}\nabla_r\varphi=\frac{\hbar}{m}\left(\frac{\ell}{r\sin\theta}\vec{e}_\phi+\frac{\tau}{r}\vec{e}_\theta\right). \tag{1.17}$$

It might seem that the form of velocity (1.17) has a purely artificial mathematical nature, but this is not the case. Let us illustrate the physical meaning of the quantity $\langle\vec{v}\rangle^{(\ell,\tau)}$, by calculating its circulation along a closed contour $\Gamma$ with length element $d\vec{l}=\vec{e}_\theta r d\theta+\vec{e}_\phi r\sin\theta d\phi$ [43]

$$\oint_\Gamma \langle\vec{p}\rangle^{(\ell,\tau)} d\vec{l}=2\pi\hbar\ell=h\ell, \tag{1.18}$$

where $2\pi\hbar=h$ is the Planck constant, and $\langle\vec{p}\rangle^{(\ell,\tau)}=m\langle\vec{v}\rangle^{(\ell,\tau)}$ is the average flux of the momentum field. For $\ell\in\mathbb{Z}$ expression (1.18) is the Bohr-Sommerfeld quantization principle [48]. The presence of a pole on the OZ axis leads to a non-zero right-hand side in (1.18). The radius $\rho=r\sin\theta$ of the region around the pole, where the probability flux velocity (1.17) can exceed the speed of light, is determined by the Compton wavelength $\lambda\!\!\!^{-}_C$. Indeed, when $\langle\vec{v}\rangle^{(\ell,\tau)}=c=\hbar\ell/m\rho$, we obtain $\rho_\ell=\frac{\hbar\ell}{mc}=\lambda\!\!\!^{-}_C\ell$.

Note that the field (1.17) can be represented in another way by explicitly isolating the vortex component $\vec{A}$

$$\varphi'(\theta,\phi,t)=\tau\theta-\frac{\mathrm{E}}{\hbar}t,\quad \vec{A}=-\frac{\hbar\ell}{qr\sin\theta}\vec{e}_\phi, \tag{1.19}$$

$$\langle\vec{v}\rangle^{(\ell,\tau)}(r,\theta)=\frac{\hbar}{m}\nabla_r\varphi'-\frac{q}{m}\vec{A}, \tag{1.20}$$

where $\varphi=\varphi'+\tilde{\varphi}$ (i.5). In both cases (1.17) and (1.20), the same expression is obtained. Figure 1 shows the lines of the vector field $\langle\vec{v}\rangle^{(\ell,\tau)}$ on concentric spheres for $\ell=4$, $\tau=1$. The probability flux has a vortex structure and flows from one pole of the sphere to the other. On the axis connecting the poles of all concentric spheres, the field (1.17) has a singularity. The current lines at the poles do not intersect and converge to the pole infinitely long. In [43] an astrophysical illustration of field (1.17) was proposed. The pole into which the current lines enter was conventionally called a «black hole», and the pole from which the current lines emerge – a «white hole». The axis connecting the poles and having a singularity was called a «naked singularity» or «wormhole». The dissipation sources $Q_1$ (1.3) for the field $\langle\vec{v}\rangle^{(\ell,\tau)}$ have the form

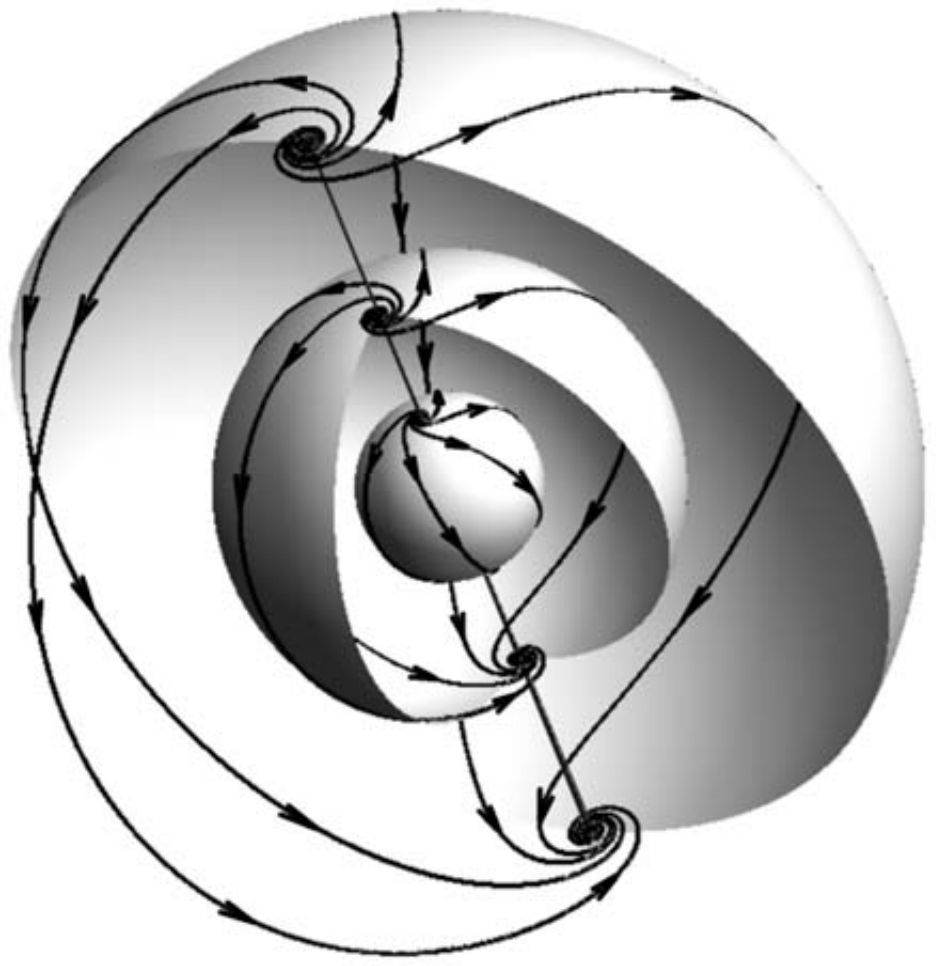

Fig. 1 Probability flux $\langle\vec{v}\rangle^{(4,1)}$

$$Q_1^{(\tau)}(r,\theta)=\frac{\hbar\tau}{mr^2}\operatorname{ctg}\theta. \tag{1.21}$$

Thus, on the equator (see Fig. 1, $\theta=\pi/2$) dissipation sources are absent. In the upper hemisphere ($0<\theta<\pi/2$) the density is positive $Q_1^{(\tau)} > 0$, and in the lower hemisphere ($\pi/2 < \theta < \pi$) the density is negative $Q_1^{(\tau)} < 0$. The «white hole» emits dissipation, while the «black hole» absorbs dissipation. In the «wormhole» (on the axis), the density of dissipation sources is unbounded. Moving away from the axis, the density of dissipation sources decreases as $Q_1^{(\tau)} \sim 1/r^2$ (1.21). According to the Vlasov equation (1.1), the sign $-Q$ determines the change in probability density along phase trajectories. As a result, near the «black hole» $-Q_1^{(\tau)} > 0$ and the probability density increases, while near the «white hole» $-Q_1^{(\tau)} < 0$, so the probability density decreases. Essentially, there is a redistribution of probability density from one pole to another over the sphere's surface (see Fig. 1) and the connection of these fluxes through the singular axis («wormhole»). This process occurs on each nested sphere connected by a common «wormhole».

The solution to the first Vlasov equation (1.1) with field (1.17) can be found by the method of characteristics ($\xi,\eta$) and has the form [43]

$$\xi(\theta,\phi) = \phi + \frac{\ell}{\tau}\operatorname{ctg}\theta, \quad \eta(r,\theta,t) = \theta - \omega_\tau(r)t, \tag{1.22}$$

$$f_1^{(\ell,\tau)}(\vec{r},t) = \frac{1}{\sin\theta}F_0\left[r, \phi + \frac{\ell}{\tau}\left(\operatorname{ctg}\theta - \operatorname{ctg}\left[\theta - \omega_\tau(r)t\right]\right), \theta - \omega_\tau(r)t\right], \quad \omega_\tau(r) = \frac{\hbar\tau}{mr^2}, \tag{1.23}$$

where $F_0$ is some function determined by initial conditions. Thus, the solution to the time dependent Schrödinger equation (1.11) by virtue of (1.23) and (1.19) has the form

$$\Psi^{(\ell,\tau)}(\vec{r},t) = \pm\sqrt{f_1^{(\ell,\tau)}(\vec{r},t)}\exp\left[i\left(\tau\theta - \frac{\mathrm{E}}{\hbar}t\right)\right]. \tag{1.24}$$

According to the Hamilton-Jacobi equation (1.12) and (1.17), (1.19), the system's energy $\mathrm{E}$ can be represented as

$$\mathrm{E} = \frac{m}{2}\left|\langle\vec{v}\rangle\right|^2 + \mathrm{V} = \frac{P_\theta^2}{2m} + \frac{P_\phi^2}{2m} + \mathrm{V} = \frac{M_\phi^2}{2m\rho^2} + \frac{M_\theta^2}{2mr^2} + \mathrm{V}, \tag{1.25}$$

$$M_\theta = \hbar\tau, \qquad M_\phi = \hbar\ell, \qquad \rho = r\sin\theta.$$

According to (1.25), the kinetic energy is the sum of rotational energies along the axial angle $\theta$ and azimuthal angle $\phi$. The expressions $M_\theta$ and $M_\phi$ correspond to angular momenta.

The magnetic field $\vec{B}$ (1.14) of the system is determined by the vector potential (1.19)

$$\vec{B} = -\frac{q_m^{(Wb)}}{2\pi}\delta'(r\sin\theta)\vec{e}_z, \quad q_m^{(Wb)}q_e = 2\pi\hbar\ell, \tag{1.26}$$

where $q_m^{(Wb)}$ corresponds to the magnetic charge [38], and $q_e$ is the electron charge. The quantity $\delta'$ is the derivative of the Dirac $\delta$- function. The charge quantization condition for $q_m^{(Wb)}$ indicates integer values of the parameter $\ell$, related to azimuthal rotation (1.16). Note that field (1.26) satisfies Maxwell's equation $\operatorname{div}_r \vec{B} = 0$ and is localized on the «wormhole». Expressions (1.19) and (1.26) define a quantum system for which the Aharonov–Bohm effect can be observed [54-57].

## §2 Singular solutions

Using the material from §1, we construct an exact solution to the Schrödinger equation (1.11). Note that the representations of the wave function's phase (1.16) and (1.19) are related by gauge invariance (i.5). In case (1.16), the magnetic field is absent, while in (1.19) it is present. Both options (1.16) and (1.19) correspond to the same physical system with probability flux $\langle \vec{v} \rangle^{\ell,\tau}$.

According to the $\Psi$ - the time independent solution to the first Vlasov equation can be represented in the form (1.24)

$$f(r,\theta,\phi) = R^2(r)\frac{\Upsilon^2(\xi)}{\sin\theta}, \tag{2.1}$$

where $R, \Upsilon$ are some functions to be determined. Consequently, the wave function (1.24)

$$\Psi(r,\theta,\phi,t) = \frac{R(r)}{\sqrt{\sin\theta}}\Upsilon\left(\phi + \frac{\ell}{\tau}\operatorname{ctg}\theta\right)e^{i\left(\ell\phi+\tau\theta-\frac{\mathrm{E}}{\hbar}t\right)}. \tag{2.2}$$

will be a solution to the Schrödinger equation

$$i\hbar\frac{\partial\Psi}{\partial t} = -\frac{\hbar^2}{2m}\Delta_r\Psi + U\Psi. \tag{2.3}$$

As a result, the following theorem holds.

**Theorem 1** *For fixed $\ell \in \mathbb{Z}$ and $\kappa \in \mathbb{N}_0 = \mathbb{N} \cup \{0\}$ , the wave functions*

$$\Psi_n^{(\kappa,\ell,\pm\tau)}(\vec{r},t) \stackrel{\text{def}}{=} \psi_n^{(\kappa,\ell)}(r,\theta,\phi,t) = \frac{N_n^{(\kappa)} r^\kappa}{\sqrt{\sin\theta}} L_n^{(\kappa+1/2)}\left(2\nu r^2\right) e^{-\nu r^2 + i\left(\ell\phi \pm \tau\theta - \frac{\mathrm{E}_n^{(\kappa)}}{\hbar}t\right)}, \tag{2.4}$$

*where $L_n^{(\kappa+1/2)}$ are generalized Laguerre polynomials, $n \in \mathbb{N}_0$, $\pm\tau = \kappa + 1$,*

$$\mathrm{E}_n^{(\kappa)} = \frac{\hbar^2}{2m\sigma_r^2}\left(2n + \kappa + \frac{3}{2}\right), \quad N_n^{(\kappa)} = \frac{2^n}{\pi\sigma_r^{\kappa+1}}\sqrt{\frac{2^\kappa n!(n+\kappa)!}{\sigma_r\sqrt{2\pi}(2n+2\kappa+1)!}}, \quad \nu = \frac{1}{4\sigma_r^2},$$

*are solutions to the Schrödinger equation (2.3) with potential*

$$U^{(\kappa,\ell)}(r,\theta) = \frac{\hbar^2}{8m\sigma_r^2}\left[\frac{r^2}{\sigma_r^2} - (4\kappa+3)\frac{\sigma_r^2}{r^2} + \left(1-4\ell^2\right)\frac{\sigma_r^2}{r^2\sin^2\theta}\right], \tag{2.5}$$

*and quantum potential*

$$\mathrm{Q}_n^{(\kappa)}(r,\theta) = -\frac{\hbar^2}{8m\sigma_r^2}\left\{\frac{r^2}{\sigma_r^2} + \left[4\kappa(\kappa+1)+1\right]\frac{\sigma_r^2}{r^2} + \frac{\sigma_r^2}{r^2\sin^2\theta} - 8n - 4\kappa - 6\right\}, \tag{2.6}$$

*where $\sigma_r$ is a constant value corresponding to the characteristic size of the system.*

The proof of Theorem 1 is given in Appendix A.

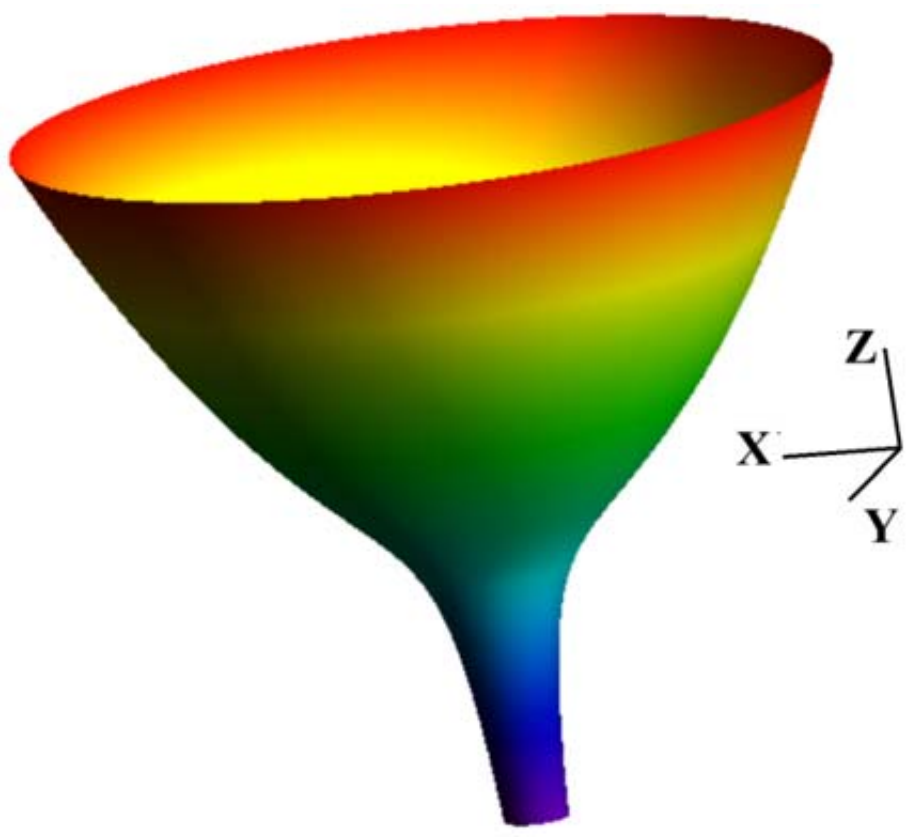

Fig. 2 Potential «funnel»

Figure 2 shows a plot of potential (2.5) in the form of a so-called «quadratic funnel». A special case of potential (2.5) for $\kappa=-3/4$ was considered previously in [24]. For $\kappa=-3/4$ and $\ell=\pm 1/2$ expression (2.5) formally transforms into the harmonic oscillator potential. Note that the energy eigenvalues $\mathrm{E}_n^{(\kappa)}$ (2.4) also coincide with the energies of the harmonic oscillator. Unlike the harmonic oscillator, the considered system has a non-zero ($\tau\neq 0$) probability flux velocity (1.17), leading to wave functions (2.4). The singularity of the wave function $1/\sqrt{\sin\theta}$ does not affect the fulfillment of the normalization condition $N_n^{(\kappa)}$ (2.4).

The quantities (1.17), (2.5), and (2.6) satisfy the Hamilton-Jacobi equation (1.25),

$$\mathrm{E}_n^{(\kappa)}=\mathrm{T}^{(\kappa,\ell)}+\mathrm{V}_n^{(\kappa,\ell)}, \tag{2.7}$$

where the kinetic $\mathrm{T}^{(\kappa,\ell)}$ and potential $\mathrm{V}_n^{(\kappa,\ell)}$ energies have the form

$$\mathrm{T}^{(\kappa,\ell)}=\frac{m}{2}\left|\langle\vec{v}\rangle^{(\ell,\tau)}\right|^2=\frac{\hbar^2}{2m}\left(\frac{\ell^2}{r^2\sin^2\theta}+\frac{(\kappa+1)^2}{r^2}\right), \tag{2.8}$$

$$\mathrm{V}_n^{(\kappa,\ell)}(r,\theta)=U^{(\kappa,\ell)}+\mathrm{Q}_n^{(\kappa)}=\frac{\hbar^2}{2m\sigma_r^2}\left[-(\kappa+1)^2\frac{\sigma_r^2}{r^2}-\frac{\ell^2\sigma_r^2}{r^2\sin^2\theta}+2n+\kappa+\frac{3}{2}\right].$$

The equations of motion (1.7) and (1.13) take the form

$$m\frac{d}{dt}\langle\vec{v}\rangle^{(\ell,\tau)}=-\nabla_r\mathrm{V}_n^{(\kappa,\ell)},\qquad \frac{1}{m}\frac{\partial\mathrm{Q}_n^{(\kappa)}}{\partial x^s}=\frac{1}{f_1}\frac{\partial P_{\lambda s}}{\partial x^\lambda}. \tag{2.9}$$

The wave function (2.4) corresponds to a vortex field of probability flux (1.19). The vortex direction is around the axial $OZ$, i.e., along the azimuthal angle $\phi$. At the same time, the probability density $f_n$ does not depend on $\phi$

$$f_n(r,\theta)=\left[N_n^{(\kappa)}\right]^2 r^{2s}\left[L_n^{(s+1/2)}\left(2\nu r^2\right)\right]^2\frac{e^{-2\nu r^2}}{\sin\theta}. \tag{2.10}$$

Figure 3 (left) shows the distribution (2.10) in the plane $\theta=\pi/2$ for $\kappa=1$, $n=3$. The sign $\operatorname{sgn}\ell$ determines the direction of the vortex part of the probability flux (1.17), and $\operatorname{sgn}\ell$ does not affect the form of the potential itself (2.5). The red and blue arrows in Fig. 3 (left) show possible directions of motion of the probability flux (1.17). The flux $\frac{\hbar\ell}{mr\sin\theta}\vec{e}_\phi$ in Fig. 3 (left) has azimuthal symmetry and depends on the polar radius $\rho=r\sin\theta$.

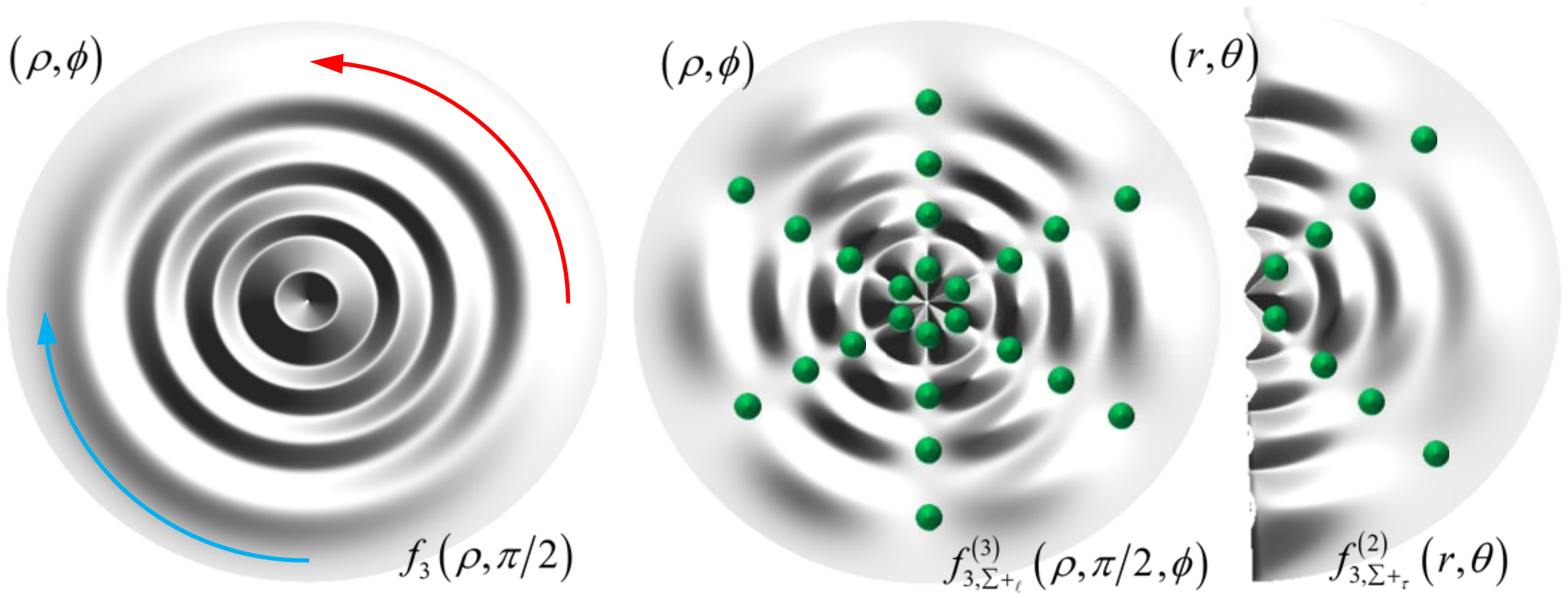

Fig. 3 Probability density distribution

Let us consider the superposition of two solutions (2.4) with opposite vortex fluxes $\Psi_n^{(\kappa,\pm\ell,\tau)}$ (see Fig. 4)

$$\Psi_{n,\Sigma_\ell}^{(\kappa,\ell,\tau)}(\vec{r},t)=c_1\Psi_n^{(\kappa,+\ell,\tau)}(\vec{r},t)+c_2\Psi_n^{(\kappa,-\ell,\tau)}(\vec{r},t)=\left(c_1e^{i\ell\phi}+c_2e^{-i\ell\phi}\right)\psi_n^{(\kappa,0,\tau)}(r,\theta,t), \quad (2.11)$$

$$\psi_n^{(\kappa,0,\tau)}(r,\theta,t)=\frac{N_n^{(\kappa)}r^\kappa}{\sqrt{\sin\theta}}L_n^{(\kappa+1/2)}\left(2\nu r^2\right)e^{-\nu r^2+i\left(\tau\theta-\frac{\mathrm{E}_n^{(\kappa)}}{\hbar}t\right)},$$

where the function $\psi_n^{(\kappa,0,\tau)}(r,\theta,t)$ has a vortex-free velocity field $\langle\vec{v}\rangle^{(0,\tau)}=\dfrac{\hbar\tau}{mr}\vec{e}_\theta$, a $c_1,c_2$ are expansion coefficients. Note that the normalization condition for the wave function (2.11) and the requirement of its continuity in the azimuthal angle $\phi$ lead to values $\ell\in\mathbb{Z}$ (Appendix B) and $|c_1|^2+|c_2|^2=1$. If $c_2=\pm c_1=\pm 1/\sqrt{2}$, then the wave function (2.11) also has a vortex-free field $\langle\vec{v}\rangle^{(0,\tau)}$

$$\Psi_{n,\Sigma\pm_\ell}^{(\kappa,\ell,\tau)}(\vec{r},t)=\sqrt{2}\psi_n^{(\kappa,0,\tau)}(r,\theta,t)\begin{cases}\cos(\ell\phi), \ "+",\\ i\sin(\ell\phi), \ "-".\end{cases} \quad (2.12)$$

Thus, two opposite vortices $\dfrac{\pm\hbar\ell\vec{e}_\phi}{mr\sin\theta}$ compensate each other and create a vortex-free field for the wave function (2.12) (see Fig. 4), whose probability density $f_{n,\Sigma\pm_\ell}^{(\ell)}=\left|\Psi_{n,\Sigma\pm_\ell}^{(\kappa,\ell,\tau)}\right|^2$ explicitly depends on the azimuthal angle $\phi$

$$f_{n,\Sigma\pm_\ell}^{(\ell)}(r,\theta,\phi)=2\left[N_n^{(\kappa)}\right]^2r^{2\kappa}\left[L_n^{(\kappa+1/2)}\left(2\nu r^2\right)\right]^2\frac{e^{-2\nu r^2}}{\sin\theta}\begin{cases}\cos^2(\ell\phi), \ "+",\\ \sin^2(\ell\phi), \ "-".\end{cases} \quad (2.13)$$

Figure 3 (center) shows the distribution (2.13) for the «+» and $\ell=3$ with explicit azimuthal dependence. From a physical point of view, the following interpretation is possible. Counter-directed wave fluxes $\langle\vec{v}\rangle^{(\pm\ell,0)}$ (1.17), moving clockwise and counterclockwise around the axial $OZ$, axis, as a result of superposition («interference») create an analogue of a standing wave for the probability density (2.13). The number of nodal points (green dots in Fig. 3) along the azimuth does not depend on the state number $n$ and is determined by the parameter $\ell$, i.e., $2\ell = 6$.

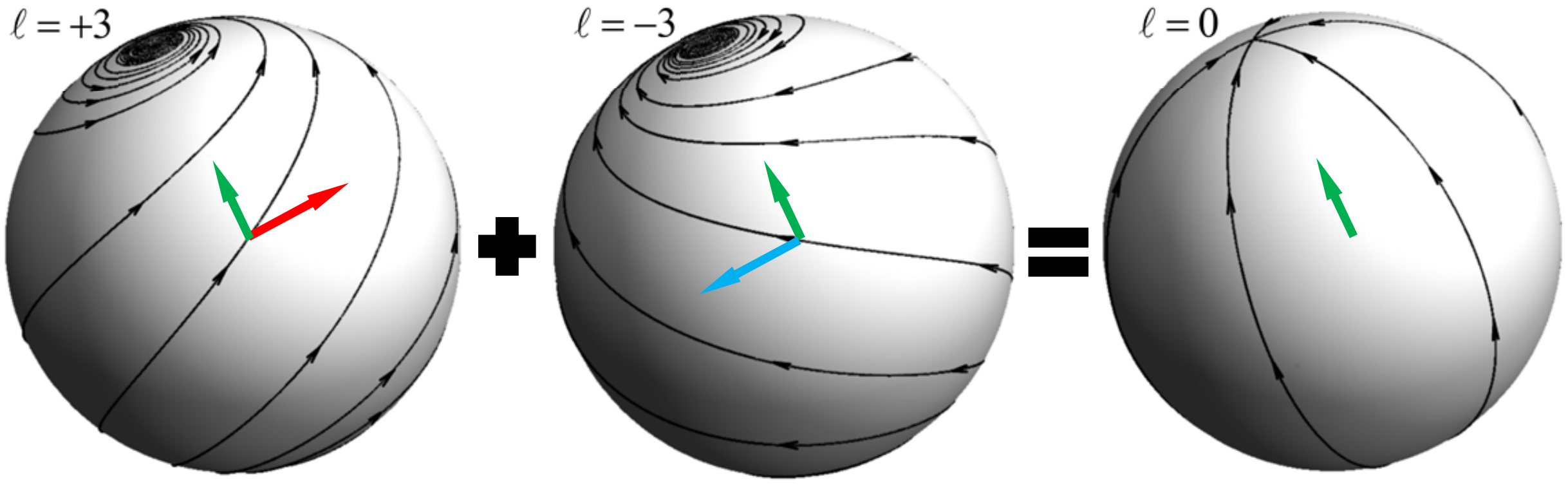

Fig. 4 Azimuthal superposition of probability currents ($\ell = \pm 3,\ \tau = -2$)

In the direction of the angle $\theta$ there is a potential flux field (1.17) $\frac{\hbar\tau}{mr}\vec{e}_\theta$, which flows from one pole of the sphere to the other (see Fig. 1) depending on the sign $\operatorname{sgn}\tau$ (2.4). For $\operatorname{sgn}\tau = +1$ flux moves from the upper pole («white hole») to the lower pole («black hole»), and for $\operatorname{sgn}\tau = -1$ from the «black hole» to the «white hole» (see Figs. 4, 5).

The potentials (2.5) and (2.6) do not depend on the direction of this flux. The value $\tau$ according to Theorem 1 depends on the parameter $\kappa$ in the potential $U^{(\kappa,\ell)}$ (2.5). The increase in the probability density (2.10) near the poles is associated with the behavior of characteristics (1.22) (see Fig. 1) and is determined by the angle $\theta$.

Note that in the presence of an axial flux $\tau \neq 0$ dissipation sources $Q_1^{(\tau)}$ (1.21) are present in the system. As a result, the probability density (2.13) is not constant (1.1) along the axial characteristics shown in Fig. 4 (right). Since for a time independent system the Boltzmann $H_n$-functions (1.6) are constant, the average dissipation sources $\langle Q_1 \rangle$ must be zero. Indeed

$$\left\langle Q_1^{(\tau)} \right\rangle_n = 2\left[N_n^{(\kappa)}\right]^2 \frac{\hbar\tau}{m} \int_0^{2\pi} \begin{cases} \cos^2(\ell\phi), & "+", \\ \sin^2(\ell\phi), & "-". \end{cases} d\phi \int_0^{\pi} \operatorname{ctg}\theta d\theta \int_0^{+\infty} r^{2\kappa} \left[L_n^{(\kappa+1/2)}\left(2\nu r^2\right)\right]^2 e^{-2\nu r^2} dr = 0, \quad (2.14)$$

where it is taken into account that $\int_0^{\pi} \operatorname{ctg}\theta d\theta = 0$.

Let us find the intrinsic magnetic moment of the system $\vec{\mu}_S$, determined by the probability density $f_1$ and the velocity field $\langle \vec{v} \rangle$

$$\vec{\mu}_S = \frac{1}{2}\int_{\mathbb{R}^3} \left(\vec{r} \times \vec{J}\right) d^3 r, \quad \vec{J} = q_e f_1 \langle \vec{v} \rangle = 2\mu_B \left(\frac{\tau}{r}\vec{e}_\theta + \frac{\ell}{r\sin\theta}\vec{e}_\phi\right) f_1, \quad (2.15)$$

where $\mu_B = q_e\hbar/2m$ is the Bohr magneton. Substituting (2.10) for $f_1$ into (2.15), we obtain (Appendix B)

$$\vec{\mu}_{\rm S}^{(1)} = \mu_{\rm B}\ell\vec{e}_z. \qquad (2.16)$$

The presence of a magnetic moment in the quantum system (2.4) is due to the vortex field (1.17), shown in Fig. 1. If the phase $\varphi = -\mathrm{E}t/\hbar$, then there is no magnetic moment.

For the system described by superposition (2.12), there is only a potential field $\langle\vec{v}\rangle^{(0,\tau)} = \dfrac{\hbar\tau}{mr}\vec{e}_\theta$ (see Fig. 4, right). As a result, the intrinsic magnetic moment (Appendix B)

$$\vec{\mu}_{\rm S}^{(2)} = \vec{\theta}. \qquad (2.17)$$

Figure 4 clearly illustrates expression (2.17). The superposition of quantum systems with magnetic moments $\pm\mu_{\rm B}\ell\vec{e}_z$, results in a system with moment: $\vec{\mu}_{\rm S}^{(2)} = +\mu_{\rm B}\ell\vec{e}_z - \mu_{\rm B}\ell\vec{e}_z = \vec{\theta}$. Note that the condition $\ell\in\mathbb{Z}$ indicates the quantization of the intrinsic magnetic moment $\vec{\mu}_{\rm S}$.

Consider the superposition of two wave functions with opposite probability fluxes $\pm\dfrac{\hbar\ell}{mr}\vec{e}_\theta$ between the sphere's poles (see Fig. 5)

$$\Psi_{n,\Sigma_\tau}^{(\kappa,\ell,\tau)}(\vec{r},t) = c_3\Psi_n^{(\kappa,\ell,+\tau)}(\vec{r},t) + c_4\Psi_n^{(\kappa,\ell,-\tau)}(\vec{r},t) = \left(c_3e^{i\tau\theta} + c_4e^{-i\tau\theta}\right)\psi_n^{(\kappa,\ell,0)}(r,\theta,\phi,t), \qquad (2.18)$$

$$\psi_n^{(\kappa,\ell,0)}(r,\theta,\phi,t) = \frac{N_n^{(\kappa)}r^\kappa}{\sqrt{\sin\theta}}L_n^{(\kappa+1/2)}\left(2\nu r^2\right)e^{-\nu r^2 + i\left(\ell\phi - \frac{\mathrm{E}_n^{(\kappa)}}{\hbar}t\right)},$$

where the function $\psi_n^{(\kappa,\ell,0)}(r,\theta,\phi,t)$ lacks the potential flux (along angle $\theta$), but contains the azimuthal vortex flux (along angle $\phi$). The normalization condition and $\theta$-symmetry require $\tau\in\mathbb{Z}$ (see Theorem 1) and $|c_3|^2 + |c_4|^2 = 1$. For $c_4 = \pm c_3 = \pm 1/\sqrt{2}$, we obtain

$$\Psi_{n,\Sigma\pm_\tau}^{(\kappa,\ell,\tau)}(\vec{r},t) = \sqrt{2}\psi_n^{(\kappa,\ell,0)}(r,\theta,\phi,t)\begin{cases}\cos(\tau\theta), \text{ "+"},\\ i\sin(\tau\theta), \text{ "−"},\end{cases} \qquad (2.19)$$

$$f_{n,\Sigma\pm_\tau}^{(\tau)}(r,\theta) = 2\left[N_n^{(\kappa)}\right]^2 r^{2\kappa}\left[L_n^{(\kappa+1/2)}\left(2\nu r^2\right)\right]^2\frac{e^{-2\nu r^2}}{\sin\theta}\begin{cases}\cos^2(\tau\theta), \text{ "+"},\\ \sin^2(\tau\theta), \text{ "−"}.\end{cases} \qquad (2.20)$$

From Eqs. (2.19)-(2.20) it follows that counter-propagating probability fluxes $\pm\dfrac{\hbar\tau}{mr}\vec{e}_\theta$ along angle $\theta$ (Fig. 5) compensate each other, leaving only the azimuthal vortex flux $\dfrac{\hbar\ell\vec{e}_\phi}{mr\sin\theta}$. The counter-propagating fluxes $\pm\dfrac{\hbar\tau}{mr}\vec{e}_\theta$ create a standing wave analog on the sphere's surface between its poles. Fig. 3 (right) shows distribution (2.20) for the «+» case with $\kappa = 1$ in the axial plane $(r,\theta)$. The number of nodal points (green dots) between the sphere's poles at each radius is determined by $2\tau = 4$ ($\tau = 2$ on each side).

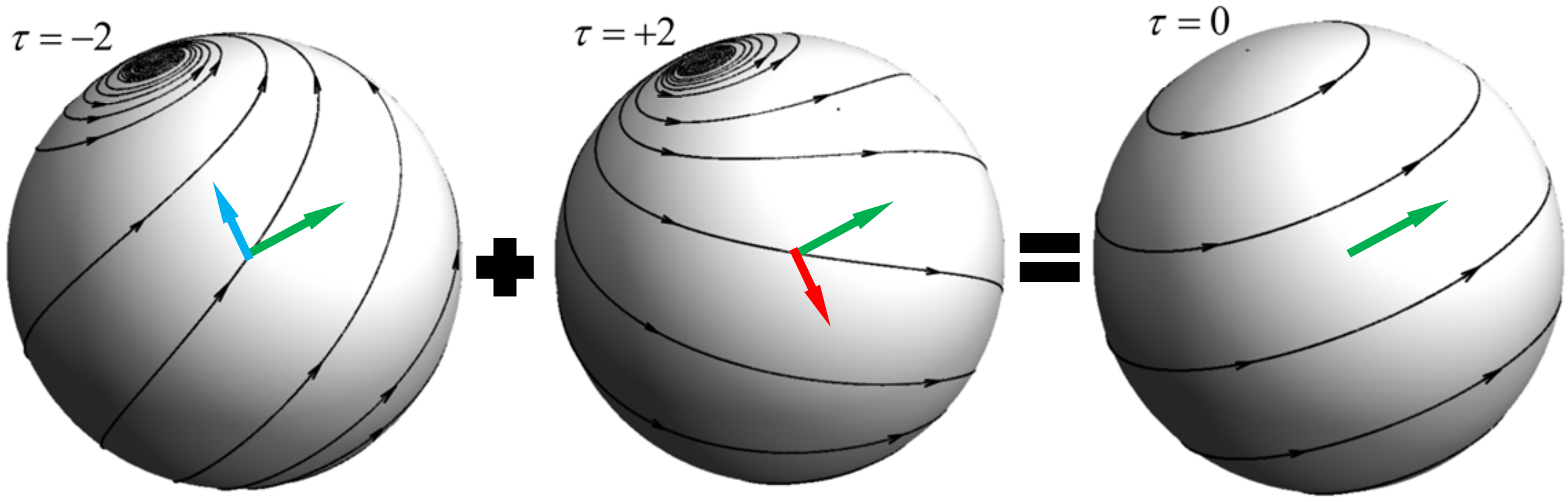

Fig. 5 Axial superposition of probability currents ( $\tau=\pm 2,\ \ell=+3$ )

The absence of axial flux $\tau=0$ (Fig. 5 right) eliminates dissipation sources $Q_1^{(\tau)}=0$. Consequently, the probability density $f_{n,\Sigma\pm_\tau}^{(\tau)}(r,\theta)$ according to the Vlasov equation (1.1), it must remain constant along the characteristics shown in Fig. 5 (right). Indeed, expression (2.20) shows that the condition is satisfied on the characteristics $\theta=const$ holds $f_{n,\Sigma\pm_\tau}^{(\tau)}=const$.

The intrinsic magnetic moment for superposition (2.19) takes the form (Appendix B)

$$\vec{\mu}_S^{(3)}=\mu_B\ell\vec{e}_z, \tag{2.21}$$

corresponding to the vortex field in Fig. 5 (right).

Similarly, one can construct superpositions from wave functions (2.12) or (2.19) with zero probability flux $\langle\vec{v}\rangle=0$. Specifically, superposing $\Psi_{n,\Sigma\pm_\ell}^{(\kappa,\ell,+\tau)}$ и $\Psi_{n,\Sigma\pm_\ell}^{(\kappa,\ell,-\tau)}$ compensates the axial flux $\frac{\hbar\tau}{mr}\vec{e}_\theta$, (Fig. 4 right), and superposing $\Psi_{n,\Sigma\pm_\tau}^{(\kappa,+\ell,\tau)}$ и $\Psi_{n,\Sigma\pm_\tau}^{(\kappa,-\ell,\tau)}$ compensates the azimuthal flux $\frac{\hbar\ell\vec{e}_\phi}{mr\sin\theta}$, (Fig. 5 right), respectively. Both combinations $\Psi_{n,\Sigma\pm_\ell}^{(\kappa,\ell,\pm\tau)}$ or $\Psi_{n,\Sigma\pm_\ell}^{(\kappa,\pm\ell,\tau)}$ yield the same wave function $\Psi_{n,\Sigma\pm_\tau,\Sigma_\ell}^{(\kappa,\ell,\tau)}$

$$\Psi_{n,\Sigma\pm_\tau,\Sigma_\ell}^{(\kappa,\ell,\tau)}(\vec{r},t)=c_5\Psi_{n,\Sigma\pm_\tau,}^{(\kappa,+\ell,\tau)}(\vec{r},t)+c_6\Psi_{n,\Sigma\pm_\tau,}^{(\kappa,-\ell,\tau)}(\vec{r},t). \tag{2.22}$$

For the special case $c_6=\pm c_5=\pm 1/\sqrt{2}$, where $|c_5|^2+|c_6|^2=1$, Eq. (2.22) becomes

$$\Psi_{n,\Sigma\pm_\tau,\Sigma\pm_\ell}^{(\kappa,\ell,\tau)}(\vec{r},t)=2\psi_n^{(\kappa,0,0)}(r,\theta,t)\begin{cases}\cos(\ell\phi)\cos(\tau\theta), \text{"+"},\\ -\sin(\ell\phi)\sin(\tau\theta), \text{"−"},\end{cases} \tag{2.23}$$

$$\psi_n^{(\kappa,0,0)}(r,\theta,t)=\frac{N_n^{(\kappa)}r^\kappa}{\sqrt{\sin\theta}}L_n^{(\kappa+1/2)}\left(2\nu r^2\right)e^{-\nu r^2-i\frac{E_n^{(\kappa)}}{\hbar}t}.$$

The probability density $f_{n,\Sigma\pm_\tau,\Sigma\pm_\ell}^{(\ell,\tau)}=\left|\Psi_{n,\Sigma\pm_\tau,\Sigma\pm_\ell}^{(\kappa,\ell,\tau)}\right|^2$ exhibits standing waves in both axial and azimuthal directions, with nodal counts determined by $\tau$ and $\ell$ (Fig. 3 center and right)

$$f_{n,\Sigma\pm_\tau,\Sigma\pm_\ell}^{(\ell,\tau)} = 4\frac{N_n^{(\kappa)}r^{2\kappa}}{\sin\theta}\left[L_n^{(\kappa+1/2)}\left(2\nu r^2\right)\right]^2 e^{-2\nu r^2}\begin{cases}\cos^2(\ell\phi)\cos^2(\tau\theta), \text{"+"},\\ \sin^2(\ell\phi)\sin^2(\tau\theta), \text{"−"}.\end{cases} \tag{2.24}$$

The absence of flux $\langle\vec{v}\rangle = 0$ leads to zero dissipation sources $Q_1 = 0$, with vanishing characteristics since no flux exists along them. Naturally, such a system's magnetic moment is zero: $\vec{\mu}_S^{(4)} = \vec{\theta}$.

Now consider the electromagnetic Schrödinger equation with vector potential (1.19)

$$i\hbar\frac{\partial\Psi}{\partial t} = \frac{1}{2m}\left(\hat{p} - q\vec{A}\right)^2\Psi + U\Psi. \tag{2.25}$$

**Theorem 2** *For fixed* $\kappa \in \mathbb{N}_0$*, the wave functions*

$$\Psi_n^{(\kappa,0,\tau)}(\vec{r},t) = \psi_n^{(\kappa,0,\tau)}(r,\theta,t) = \frac{N_n^{(\kappa)}r^\kappa}{\sqrt{\sin\theta}}L_n^{(\kappa+1/2)}\left(2\nu r^2\right)e^{-\nu r^2 + i\left(\tau\theta - \frac{E_n^{(\kappa)}}{\hbar}t\right)}, \tag{2.26}$$

*are solutions of the electromagnetic Schrödinger equation (2.25) with vector potential (1.19), scalar potential (2.5), and quantum potential (2.6).*

The proof of Theorem 2 is given in Appendix A.

The wave functions (2.26) possess the vector field $\langle\vec{v}\rangle^{(0,\tau)} = \frac{\hbar\tau}{mr}\vec{e}_\theta$, shown in Fig. 4 (right). Due to dissipation sources $Q_1^{(\tau)}$, the corresponding probability density $\left|\Psi_n^{(\kappa,0,\tau)}\right|^2$ varies along $\langle\vec{v}\rangle^{(0,\tau)}$ matching $f_n(r,\theta)$ (2.10). consequently, the intrinsic magnetic moment vanishes: $\vec{\mu}_S^{(5)} = \vec{\theta}$. Indeed, the vortex field is now contained in the external magnetic field via vector potential (1.19).

A zero field $\langle\vec{v}\rangle = \vec{\theta}$ can be obtained by superposing $\Psi_n^{(\kappa,0,\pm\tau)}$. Analogously to previous cases

$$\Psi_{n,\Sigma\pm_\tau}^{(\kappa,0,\tau)}(\vec{r},t) = \sqrt{2}\psi_n^{(\kappa,0,0)}(r,\theta,t)\begin{cases}\cos(\tau\theta), \text{"+"},\\ i\sin(\tau\theta), \text{"−"},\end{cases} \tag{2.27}$$

$$\psi_n^{(\kappa,0,0)}(r,\theta,t) = \frac{N_n^{(\kappa)}r^\kappa}{\sqrt{\sin\theta}}L_n^{(\kappa+1/2)}\left(2\nu r^2\right)e^{-\nu r^2 - i\frac{E_n^{(\kappa)}}{\hbar}t}.$$

The probability density of superposition (2.27) coincides with (2.20) $f_{n,\Sigma\pm_\tau}^{(\tau)}(r,\theta)$, and the intrinsic magnetic moment is zero: $\vec{\mu}_S^{(6)} = \vec{\theta}$.

## §3 Bounded solutions

Consider the special case $\tau = 0$ of the phase (1.16). Then the dissipation sources (1.21) vanish, and the velocity field (1.17) contains only the vortex component. The first Vlasov equation (1.1) takes the form

$$-\frac{mr^2}{\hbar}\frac{\partial S}{\partial t}+\frac{\ell}{\sin^2\theta}\frac{\partial S}{\partial\phi}=0, \tag{3.1}$$

which admits a solution by the method of characteristics

$$S(r,\theta,\phi,t)=F\left[r,\theta,\varsigma(\phi,\theta,t)\right],\ \ \varsigma(\phi,\theta,t)=\phi-\frac{\hbar\ell t}{mr^2\sin^2\theta}. \tag{3.2}$$

Since we are interested in time independent probability density distributions, the solution to the Schrödinger equation (2.3) can be sought in the form

$$\Psi=R(r)\Theta(\theta)e^{i\left(\ell\phi-\frac{\mathrm{E}}{\hbar}t\right)}, \tag{3.3}$$

where $R,\Theta$ are functions to be determined.

**Theorem 3** *Let the numbers* $(l,\varepsilon,\ell)$ *be related by the Pythagorean triple relation* $l^2+\varepsilon^2=\ell^2$, *then the wave functions*

$$\Psi_{n,s}^{(\pm l,\pm\ell)}(\vec{r},t)\overset{\mathrm{def}}{=}\psi_{n,s}^{(\mp l,\pm\ell)}(r,\theta,\phi,t)=N_{n,s}^{(\pm l)}r^sL_n^{(s+1/2)}\left(2\nu r^2\right)P_s^{\pm l}(\cos\theta)e^{-\nu r^2+i\left(\pm\ell\phi-\frac{\mathrm{E}_{n,s}}{\hbar}t\right)}, \tag{3.4}$$

*where* $0\le l\le s$, $s\in\mathbb{N}_0$, $P_s^l$ *are the associated Legendre polynomials,*

$$\mathrm{E}_{n,s}=\frac{\hbar^2}{2m\sigma_r^2}\left(2n+s+\frac{3}{2}\right),\qquad N_{n,s}^{(l)}=\frac{2^n}{\sigma_r^{s+1}}\sqrt{\frac{2^{s-1}n!(n+s)!(2s+1)(s-l)!}{\sigma_r\pi\sqrt{2\pi}\,(2n+2s+1)!(s+l)!}},$$

*are solutions to the Schrödinger equation (2.3) with potential*

$$U^{(\varepsilon)}(r,\theta)=\frac{\hbar^2}{8m\sigma_r^2}\left(\frac{r^2}{\sigma_r^2}-\frac{4\varepsilon^2\sigma_r^2}{r^2\sin^2\theta}\right), \tag{3.5}$$

*and quantum potential*

$$\mathrm{Q}_{n,s}^{(l)}(r,\theta)=-\frac{\hbar^2}{8m\sigma_r^2}\left(\frac{r^2}{\sigma_r^2}+\frac{4l^2\sigma_r^2}{r^2\sin^2\theta}-8n-4s-6\right). \tag{3.6}$$

The proof of Theorem 3 is given in Appendix B.

By performing a gauge transformation of the electromagnetic field, analogous to Theorem 2, we can formulate the following theorem for a system with potential (3.5):

**Theorem 4** *Let the numbers* $(l,\varepsilon,\ell+\kappa)$ *related by the Pythagorean triple relation* $l^2+\varepsilon^2=(\ell+\kappa)^2$, *then the wave functions*

$$\Psi_{n,s}^{(\pm l,\pm\kappa)}(\vec{r},t) \overset{\text{def}}{=} \psi_{n,s}^{(\mp l,\pm\kappa)}(r,\theta,\phi,t) = N_{n,s}^{(\pm l)} r^s L_n^{(s+1/2)}(2\nu r^2) P_s^{\pm l}(\cos\theta) e^{-\nu r^2 + i\left(\kappa\phi - \frac{E_{n,s}}{\hbar}t\right)},\ 0 \le l \le s, \quad (3.7)$$

*are solutions to the electromagnetic Schrödinger equation (2.25) with vector potential (1.19), scalar potential (3.5), and quantum potential (3.6).*

The proof of Theorem 4 is given in Appendix A.

Note that Pythagorean triples are defined by two numbers $j,k \in \mathbb{N}$ of different parity as $(j^2-k^2, 2jk, j^2+k^2)$, where $j>k$. The set of Pythagorean triples has the form

$$\mathbb{N}_3 \overset{\text{def}}{=} \{(l,\varepsilon,\upsilon):(3,4,5);(5,12,13);(8,15,17);(7,24,25);....\}. \quad (3.8)$$

Thus in Theorems 3 and 4 the values $(l,\varepsilon,\ell)$ or $(l,\varepsilon,\ell+\kappa)$ can be expressed through $j,k$, but it is also allowed that $j,k \in \mathbb{N}_0$ and have the same parity.

The potential (3.5) with $\varepsilon = 1$ was considered in [24] for studying the properties of the Wigner and Weyl-Stratonovich functions. In that work, only one solution (3.4) or (3.7) was obtained for the state $n=s=l=0$ and $\ell=1$. Here, Theorems 3 and 4 present extended sets of solutions.

Note that for $\varepsilon = 0$ ($l=\ell$) the potential (3.5) corresponds to a harmonic oscillator. Consequently, the solution (3.4) transforms into the known solution for the harmonic oscillator

$$\Psi_{n,s}^{(\pm\ell,\pm\ell)}(\vec{r},t) = N_{n,s} r^s L_n^{(s+1/2)}(2\nu r^2) Y_s^{\pm\ell}(\theta,\phi) e^{-\nu r^2 - i\frac{E_{n,s}}{\hbar}t}, \quad (3.9)$$

$$N_{n,s} = \frac{2^n}{\sigma_r^{s+1}} \sqrt{\frac{2^{s+1} n!(n+s)!}{\sigma_r \sqrt{2\pi}(2n+2s+1)!}},$$

where we have used the definition of spherical harmonics

$$Y_s^{\ell}(\theta,\phi) = \sqrt{\frac{(2s+1)(s-\ell)!}{4\pi(s+\ell)!}} e^{i\ell\phi} P_s^{\ell}(\cos\theta). \quad (3.10)$$

A similar result can also be achieved for the solution (3.7) in the absence of magnetic field ($\ell = 0$) when $\varepsilon = 0$ ($l = \pm\kappa$). Therefore, the quantum system (3.4)-(3.6) represents a more general case than the harmonic oscillator. Analogous to the spherical harmonics (3.10), we can define modified spherical harmonics $Y_s^{(\ell,l)}$ for the solution (3.4)

$$Y_s^{(\ell,l)}(\theta,\phi) \overset{\text{def}}{=} \sqrt{\frac{(2s+1)(s-l)!}{4\pi(s+l)!}} e^{i\ell\phi} P_s^{l}(\cos\theta). \quad (3.11)$$

In the special case when $\ell = l$ the functions (3.11) and (3.10) coincide, i.e., $Y_s^{(\ell,\ell)} = Y_s^{(\ell)}$. Consequently, the solution (3.4) can be represented as

$$\Psi_{n,s}^{(\pm l,\pm\ell)}(\vec{r},t) = N_{n,s}^{(\pm l)} r^s L_n^{(s+1/2)}(2\nu r^2) Y_s^{(\pm l,\pm\ell)}(\theta,\phi) e^{-\nu r^2 - i\frac{E_{n,s}}{\hbar}t}. \quad (3.12)$$

For the functions (3.11), orthogonality relations similar to those for spherical harmonics (3.10) hold.

**Lemma 1** *The modified spherical harmonics (3.11) satisfy the orthogonality relations*

$$\int_{S^2} Y_s^{(\ell,l)} Y_{s'}^{(\ell',l)*} ds = \delta_{\ell,\ell'} \delta_{s,s'}, \qquad \int_{S^2} \frac{1}{\sin^2\theta} Y_s^{(\ell,l)} Y_s^{(\ell',l')*} ds = \frac{2s+1}{2l} \delta_{\ell,\ell'} \delta_{l,l'}, \tag{3.13}$$

*where integration is performed over the sphere* $S^2$ *surface,* $ds = \sin\theta d\theta d\phi$ *and for* $l = 0$ *second relation in (3.13) becomes unbounded.*

The proof of Lemma 1 is given in Appendix B.

Let us examine in more detail the combinations of Pythagorean triples $(l, \varepsilon, \ell)$ for the solution (3.4). There are three possible situations for different values of $n$ and $s$:

- $\forall \varepsilon \in \mathbb{N} : l = 0,\ \ell = \varepsilon$
- $\exists (l, \varepsilon, \ell) \in \mathbb{N}_3,\ l \leq s$ (3.14)
- $\varepsilon = 0 : l = \ell \leq s$ (harmonic oscillator)

The last case in (3.14) of the harmonic oscillator is well-studied, so we will not focus on it. Figure 6 shows examples of distributions of the real part of the function (3.11) for various Pythagorean triples $(l, \varepsilon, \ell)$ from the first two variants in (3.14). Green areas in Fig. 6 represent regions inside a transparent unit sphere, while gray areas show regions outside the unit sphere. The unit sphere conventionally corresponds to the «undeformed» state. At each point $(\theta, \phi)$ the value of $\mathrm{Re}\, Y_s^{(\ell,l)}(\theta, \phi)$ is computed.

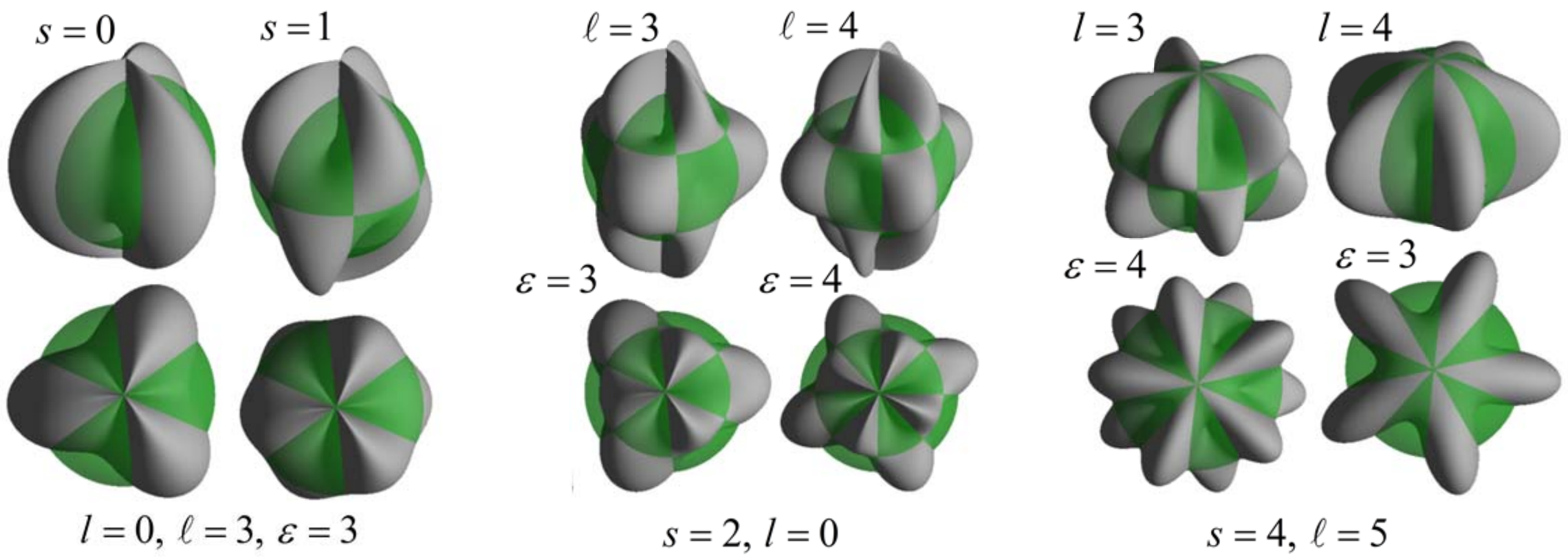

Fig. 6 Functions $\mathrm{Re}\, Y_s^{(\ell,l)}(\theta, \phi)$ for different Pythagorean triples

For example, the function $Y_0^{(0,0)}(\theta, \phi) = C_0$ corresponds to the undeformed state. As the indices $s, l, \ell$ change, the function $\mathrm{Re}\, Y_s^{(\ell,l)}$ at each point $(\theta, \phi)$ will deviate from the undeformed state either positively or negatively, falling into either the gray or green areas, respectively.

Horizontally, Fig. 6 consists of three fragments: left, center, and right. The left and right fragments are related to changes in states with $s=0,1$ and $l=3,4$. The central fragment shows changes in the probability current $\ell=3,4$. Vertically, Fig. 6 presents various projections of the $\operatorname{Re} Y_s^{(\ell,l)}$ distributions (isometric and top views). The right fragment in Fig. 6 corresponds to the second variant in (3.14), as it contains the triplets $(3,4,5)$ and $(4,3,5)$. Such triplets are rare and are elements of the set (3.8). The central and left fragments correspond to the more common first variant in (3.14).

For the quantum system (3.4), the Hamilton-Jacobi equation (1.12) is satisfied

$$\mathrm{E}_{n,s} = \mathrm{T}^{(\ell)}(r,\theta) + \mathrm{V}_{n,s}^{(\ell)}(r,\theta), \tag{3.15}$$

$$\mathrm{T}^{(\ell)} = \frac{m}{2}\left|\langle\vec{v}\rangle^{(\ell,0)}\right|^2 = \frac{\hbar^2\ell^2}{2mr^2\sin^2\theta}, \qquad \mathrm{V}_{n,s}^{(\ell)} = \frac{\hbar^2}{2m\sigma_r^2}\left(-\frac{\ell^2\sigma_r^2}{r^2\sin^2\theta} + 2n + s + \frac{3}{2}\right).$$

The kinetic energy (3.15) is determined solely by the vortex component of the probability current. The equations of motion (2.9) are also satisfied. Since $\tau=0$, the dissipation sources (1.21) vanish $Q_1^{(\tau)}=0$. Therefore, the probability density $f_{n,s}^{(\pm l)}$ remains constant along the azimuthal current $\langle\vec{v}\rangle^{(\ell,0)}$

$$f_{n,s}^{(\pm l)}(\vec{r}) = \left[N_{n,s}^{(\pm l)}\right]^2 r^{2s}\left[L_n^{(s+1/2)}\left(2\nu r^2\right)P_s^{\pm l}(\cos\theta)\right]^2 e^{-2\nu r^2}. \tag{3.16}$$

The azimuthal current $\langle\vec{v}\rangle^{(\ell,0)}$ and the density (3.16) lead to an intrinsic magnetic moment (see Appendix B):

$$\vec{\mu}_{\mathrm{S}}^{(7)} = \mu_{\mathrm{B}}\ell\vec{e}_z. \tag{3.17}$$

Analogous to §2 construct a superposition of wave functions (3.4) that yields a zero-probability current $\langle\vec{v}\rangle=\vec{\theta}$. Such a superposition should nullify the azimuthal current

$$\Psi_{n,s,\Sigma\pm_\upsilon}^{(l,\ell)}(\vec{r},t) = \sqrt{2}\psi_{n,s}^{(l,0)}(r,\theta,t)\begin{cases}\cos(\ell\phi), & \text{"+"},\\ i\sin(\ell\phi), & \text{"−"},\end{cases} \tag{3.18}$$

$$\psi_{n,s}^{(l,0)}(r,\theta,t) = N_{n,s}^{(l)} r^s L_n^{(s+1/2)}\left(2\nu r^2\right)P_s^l(\cos\theta)e^{-\nu r^2 - i\frac{\mathrm{E}_{n,s}}{\hbar}t},$$

$$f_{n,s,\Sigma\pm_\ell}^{(l,\ell)}(r,\theta,\phi) = 2\left[N_{n,s}^{(l)}\right]^2 r^{2s}\left[L_n^{(s+1/2)}\left(2\nu r^2\right)P_s^l(\cos\theta)\right]^2 e^{-2\nu r^2}\begin{cases}\cos^2(\ell\phi), & \text{"+"},\\ \sin^2(\ell\phi), & \text{"−"}.\end{cases} \tag{3.19}$$

The wave function (3.18) has zero magnetic moment $\vec{\mu}_{\mathrm{S}}^{(8)}=\theta$ and zero kinetic energy (3.15). At the same time, the function (3.18) satisfies the Schrödinger equation with potential (3.5), but the quantum potential differs from (3.6). Indeed, from the equation of motion (1.13) or (2.9) and the Hamilton-Jacobi equation it follows that

$$\vec{\theta} = -\nabla_r \mathrm{Q}_{n,s}^{(\varepsilon)} - \nabla_r U^{(\varepsilon)} \;\Rightarrow\; \nabla_r \mathrm{Q}_{n,s}^{(\varepsilon)} = -\nabla_r U^{(\varepsilon)} \;\Rightarrow\; \mathrm{Q}_{n,s}^{(\varepsilon)} = \mathrm{E}_{n,s} - U^{(\varepsilon)}, \tag{3.20}$$

$$\mathrm{Q}_{n,s}^{(\varepsilon)}(r,\theta) = -\frac{\hbar^2}{8m\sigma_r^2}\left(\frac{r^2}{\sigma_r^2} - \frac{4\varepsilon^2\sigma_r^2}{r^2\sin^2\theta} - 8n - 4s - 6\right),$$

i.e., the external force $-\nabla_r U^{(\varepsilon)}$, acting on the quantum system is balanced by the quantum pressure force. As a result, the (superposition) system (3.18) is «at rest» with the average probability current velocity $\langle\vec{v}\rangle = \vec{\theta}$. Direct calculation of the forces acting on the system (3.18) gives

$$\vec{F}_U = -\nabla_r U^{(\varepsilon)} = -\frac{\hbar^2 r}{4m\sigma_r^4}\vec{e}_r - \frac{\hbar^2\varepsilon^2}{m\rho^3}\vec{e}_\rho, \quad \vec{e}_\rho = \vec{e}_r\sin\theta + \vec{e}_\theta\cos\theta, \tag{3.21}$$

$$\vec{F}_{\mathrm{Q}} = -\nabla_r \mathrm{Q}_{n,s}^{(\varepsilon)} = -\vec{F}_U, \tag{3.22}$$

where $\vec{e}_\rho$ is the unit basis vector corresponding to the cylindrical coordinate system $(\rho,\phi,z)$. From expression (3.21), we see that the external force $\vec{F}_U$ compresses the system in the radial ( $\vec{e}_r$ and $\vec{e}_\rho$) directions in both spherical and cylindrical coordinate systems. The quantum pressure force $\vec{F}_{\mathrm{Q}}$ (3.22) is directed oppositely, counteracting the external force $\vec{F}_U$. The resultant force is zero: $\vec{F}_{\mathrm{Q}} + \vec{F}_U = -\nabla_r \mathrm{V} = \vec{\theta}$ (3.20).

In the original system (3.4), there exists an azimuthal rotational velocity $\langle\vec{v}\rangle^{(\ell,0)}$, leading to an analogue of «centripetal» acceleration (1.2)

$$\hat{\pi}_1\langle\vec{v}\rangle^{(\ell,0)} = \frac{d}{dt}\langle\vec{v}\rangle^{(\ell,0)} = \left(\langle\vec{v}\rangle^{(\ell,0)}\cdot\nabla_r\right)\langle\vec{v}\rangle^{(\ell,0)} = -\frac{\hbar^2\ell^2}{m^2\rho^3}\vec{e}_\rho. \tag{3.23}$$

Unlike the classical concept of centripetal acceleration, the acceleration (3.23) decreases with radius $\rho$ and increases when approaching the center (see §1). This feature is caused by the different dependence of the velocity $\langle\vec{v}\rangle^{(\ell,0)}$ on $\rho$. In classical mechanics, the linear velocity during circular motion increases with radius. In the case of $\langle\vec{v}\rangle^{(\ell,0)} = \frac{\hbar\ell\vec{e}_\phi}{m\rho}$ the velocity increases when approaching the center and decreases when moving away from it (see §1 (1.17)-(1.18)). This peculiarity leads to an acceleration of the form (3.23) and the Bohr-Sommerfeld quantization rule.

The appearance of acceleration (3.23) leads to an analogue of centrifugal force pushing the system apart. The external force $\vec{F}_U$ (3.21) remains unchanged, as it is determined by the fixed potential $U^{(\varepsilon)}$. Compensation of the centrifugal force is possible by weakening the quantum pressure force $\vec{F}_{\mathrm{Q}}$, but this requires a change in quantum pressure resulting from the transformation of the probability density (3.19) into (3.16). Indeed, the pressure force $\vec{F}'_{\mathrm{Q}}$ from the quantum potential (3.6) has the form

$$\vec{F}'_{\mathrm{Q}} = -\nabla_r \mathrm{Q}_{n,s}^{(l)} = \vec{F}_Q - \frac{\hbar^2\ell^2}{m\rho^3}\vec{e}_\rho, \tag{3.24}$$

which contains an additional term that weakens the pressure force $\vec{F}_Q$ (3.22). The additional term in (3.24) exactly matches the acceleration (3.23). Thus, the equation of motion (1.13), corresponding to the Hamilton-Jacobi equation (3.15), is fully satisfied:

$$m\frac{d}{dt}\langle\vec{v}\rangle^{(\ell,0)} = \vec{F}'_Q + \vec{F}_U = -\nabla_r \mathrm{V}_{n,s}^{(\ell)}. \tag{3.25}$$

A similar situation occurs in the quantum superpositions considered in §2. The presence of an intrinsic magnetic moment $\vec{\mu}_S$ is associated with the weakening of the quantum pressure force.

## Conclusion

The Wigner-Vlasov formalism enables the construction of exact analytical solutions to problems in both classical and quantum mechanics. Its connection with continuum mechanics provides an intuitive hydrodynamic interpretation of the probability current vector field, intrinsic magnetic moment, quantum pressure, and wavefunction superposition as probability wave interference. Quantum systems are described by conservation laws (see §1) familiar from classical physics, which follow directly from the Vlasov equation chain.

For the exact solutions (2.4), (2.26), (3.4), and (3.7) of the Schrödinger equation, one can derive corresponding Wigner functions (i.2) and Weyl-Stratonovich functions (i.6) satisfying the second Vlasov equation [24]. Knowledge of these quasi-probability densities $W_{n,s}(\vec{r},\vec{p})$ allows computation of mean kinematic quantities, such as the average acceleration flux $\langle\dot{v}_\lambda\rangle$ (1.15) or coordinate-dependent energy functions $\langle\mathcal{E}\rangle_{n,s}(\vec{r}) = \int W_{n,s}\mathcal{E}d^3p \big/ \int W_{n,s}d^3p$ or momentum-dependent $\langle\mathcal{E}\rangle_{n,s}(\vec{p}) = \int W_{n,s}\mathcal{E}d^3r \big/ \int W_{n,s}d^3r$, where $\mathcal{E}(\vec{r},\vec{p}) = p^2/2m + U(\vec{r})$ and $\langle\langle\mathcal{E}\rangle\rangle_{n,s} = \mathrm{E}_{n,s}$ is the energy spectrum of a quantum system. As shown in [49], these energy functions exhibit band structures in regions where the Wigner function is negative, effectively partitioning the original potential $U^{(p_1,p_2)}$ into a series of independent potentials corresponding to different quantum states. For the 3D solutions obtained here, these energy bands $\langle\mathcal{E}\rangle_{n,s}$ will be localized on concentric spherical surfaces coinciding with zeros of the Laguerre $L_n^{(s+1/2)}$ and Legendre $P_s^l$ polynomials.

In summary, this approach to constructing exact solutions of the Schrödinger and Vlasov equations aligns with methods developed in [43, 50-53].

## Appendix A

### *Proof of Theorem 1*

Expression (2.4) is obtained directly by substituting a solution of the form (2.2) into the Schrödinger equation (2.3). Let us perform the calculations step by step

$$\begin{aligned}\Delta\Psi = {} & \Upsilon\frac{e^{i\left(\ell\phi+\tau\theta-\frac{\mathrm{E}}{\hbar}t\right)}}{r\sqrt{\sin\theta}}\frac{\partial^2}{\partial r^2}(rR) + e^{i\left(\ell\phi-\frac{\mathrm{E}}{\hbar}t\right)}\frac{R}{r^2\sin\theta}\frac{\partial}{\partial\theta}\left(\sin\theta\frac{\partial}{\partial\theta}\frac{e^{i\tau\theta}}{\sqrt{\sin\theta}}\Upsilon\right) + \\ & + e^{i\left(\tau\theta-\frac{\mathrm{E}}{\hbar}t\right)}\frac{R}{r^2\sin^2\theta}\frac{1}{\sqrt{\sin\theta}}\frac{\partial^2}{\partial\phi^2}\left(\Upsilon e^{i\ell\phi}\right).\end{aligned} \tag{A.1}$$

Let us compute each term in expression (A.1)

$$\frac{\partial^2}{\partial r^2}(rR) = 2R' + rR'', \tag{A.2}$$

$$\frac{\partial}{\partial\theta}\left[\frac{e^{i\tau\theta}}{\sqrt{\sin\theta}}\Upsilon(\xi)\right] = \frac{i\tau e^{i\tau\theta}}{\sqrt{\sin\theta}}\Upsilon(\xi) - \frac{e^{i\tau\theta}\Upsilon(\xi)\cos\theta}{2\sin\theta\sqrt{\sin\theta}} - \frac{\ell e^{i\tau\theta}\Upsilon'(\xi)}{\tau\sin^2\theta\sqrt{\sin\theta}}, \tag{A.3}$$

$$\begin{aligned}
&\frac{\partial}{\partial\theta}\left[\sin\theta\frac{\partial}{\partial\theta}\frac{e^{i\tau\theta}}{\sqrt{\sin\theta}}\Upsilon(\xi)\right] = i\tau\frac{\partial}{\partial\theta}\left[\Upsilon(\xi)e^{i\tau\theta}\sqrt{\sin\theta}\right] - \frac{1}{2}\frac{\partial}{\partial\theta}\left[e^{i\tau\theta}\frac{\cos\theta}{\sqrt{\sin\theta}}\Upsilon(\xi)\right] - \\
&-\frac{\ell}{\tau}\frac{\partial}{\partial\theta}\frac{e^{i\tau\theta}\Upsilon'(\xi)}{\sin\theta\sqrt{\sin\theta}} = \left(-\frac{i\ell}{\sin\theta}\frac{\Upsilon'}{\Upsilon} - \tau^2\sin\theta + \frac{i\tau\cos\theta}{2}\right)\frac{e^{i\tau\theta}\Upsilon}{\sqrt{\sin\theta}} + \\
&+\left(\frac{\ell\cos\theta}{2\tau\sin^2\theta}\frac{\Upsilon'}{\Upsilon} - \frac{i\tau}{2}\cos\theta + \frac{\sin^2\theta+1}{4\sin\theta}\right)\frac{e^{i\tau\theta}\Upsilon}{\sqrt{\sin\theta}} + \left(\frac{3\ell\cos\theta}{2\tau\sin^2\theta}\frac{\Upsilon'}{\Upsilon} + \frac{\ell^2}{\tau^2\sin^3\theta}\frac{\Upsilon''}{\Upsilon} - \frac{i\ell}{\sin\theta}\frac{\Upsilon'}{\Upsilon}\right)\frac{e^{i\tau\theta}\Upsilon}{\sqrt{\sin\theta}} = \\
&= \left(-\frac{2i\ell}{\sin\theta}\frac{\Upsilon'}{\Upsilon} + \frac{2\ell\cos\theta}{\tau\sin^2\theta}\frac{\Upsilon'}{\Upsilon} - \tau^2\sin\theta + \frac{\sin^2\theta+1}{4\sin\theta} + \frac{\ell^2}{\tau^2\sin^3\theta}\frac{\Upsilon''}{\Upsilon}\right)\frac{e^{i\tau\theta}\Upsilon}{\sqrt{\sin\theta}},
\end{aligned} \tag{A.4}$$

$$\frac{\partial^2}{\partial\phi^2}\left[e^{i\ell\phi}\Upsilon(\xi)\right] = \frac{\partial}{\partial\phi}\left[e^{i\ell\phi}(\Upsilon' + i\ell\Upsilon)\right] = e^{i\ell\phi}\left(\Upsilon'' + 2i\ell\Upsilon' - \ell^2\Upsilon\right). \tag{A.5}$$

Substituting expressions (A.2), (A.4), and (A.5) into (A.1), we obtain

$$\frac{\Delta\Psi}{\Psi} = \frac{R''}{R} + \frac{2}{r}\frac{R'}{R} + \frac{1}{r^2\sin^2\theta}\left(\frac{\ell^2}{\tau^2\sin^2\theta} + 1\right)\frac{\Upsilon''}{\Upsilon} + \frac{2\ell\cos\theta}{\tau r^2\sin^3\theta}\frac{\Upsilon'}{\Upsilon} + \frac{1-4\ell^2}{4r^2\sin^2\theta} + \frac{1-4\tau^2}{4r^2}. \tag{A.6}$$

From the Schrödinger equation for function (2.2), it follows that the potential $U$ can be expressed as

$$\begin{aligned}
U &= \mathrm{E} + \frac{\hbar^2}{2m}\frac{\Delta\Psi}{\Psi} = \frac{\hbar^2}{8m\sigma_r^2}\left[\frac{r^2}{\sigma_r^2} - (4\kappa+3)\frac{\sigma_r^2}{r^2} + (1-4\ell^2)\frac{\sigma_r^2}{r^2\sin^2\theta}\right] = \\
&= \mathrm{E} + \frac{\hbar^2}{2m}\left[\frac{R''}{R} + \frac{2}{r}\frac{R'}{R} + \frac{1}{r^2\sin^2\theta}\left(\frac{\ell^2}{\tau^2\sin^2\theta} + 1\right)\frac{\Upsilon''}{\Upsilon} + \frac{2\ell\cos\theta}{\tau r^2\sin^3\theta}\frac{\Upsilon'}{\Upsilon} + \frac{1-4\ell^2}{4r^2\sin^2\theta} + \frac{1-4\tau^2}{4r^2}\right].
\end{aligned} \tag{A.7}$$

Comparing the right and left sides of expression (A.7), we obtain

$$\frac{r^2}{4\sigma_r^4} = \frac{2m\mathrm{E}}{\hbar^2} + \frac{R''}{R} + \frac{2}{r}\frac{R'}{R} + \frac{1+\kappa-\tau^2}{r^2} + \frac{1}{r^2\sin^2\theta}\left[\left(\frac{\ell^2}{\tau^2\sin^2\theta} + 1\right)\frac{\Upsilon''}{\Upsilon} + \frac{2\ell\cos\theta}{\tau\sin\theta}\frac{\Upsilon'}{\Upsilon}\right]. \tag{A.8}$$

The expression in square brackets (A.8) depends only on the variables $\theta$, $\phi$ and is independent of $r$, therefore

$$\left(\frac{\ell^2}{\tau^2\sin^2\theta} + 1\right)\frac{\Upsilon''}{\Upsilon} + \frac{2\ell\cos\theta}{\tau\sin\theta}\frac{\Upsilon'}{\Upsilon} = -l\sin^2\theta, \tag{A.9}$$

where $l$ is a constant. The simplest solution to equation (A.9) is $\Upsilon = const$ when $l = 0$, then, relation (A.8) becomes

$$R'' + \frac{2}{r}R' + \left(\frac{2m\mathrm{E}}{\hbar^2} + \frac{1+\kappa-\tau^2}{r^2} - \frac{r^2}{4\sigma_r^4}\right)R = 0. \quad \text{(A.10)}$$

We seek the solution to equation (A.10) in the form

$$R = r^s e^{-\nu r^2} \Lambda\left(2\nu r^2\right), \quad \text{(A.11)}$$

where $\Lambda$ is some unknown function and $\nu = const$. Substituting (A.11) into (A.10), we obtain:

$$\begin{aligned} &\frac{R'}{R} = \frac{s}{r} + 4\nu r\frac{\Lambda'}{\Lambda} - 2\nu r, \\ &\frac{R''}{R} = \frac{s(s-1)}{r^2} + 4\nu^2 r^2 + 16\nu^2 r^2\frac{\Lambda''}{\Lambda} + 4\nu\left(2s+1-4\nu r^2\right)\frac{\Lambda'}{\Lambda} - 2\nu(1+2s), \end{aligned} \quad \text{(A.12)}$$

hence:

$$\begin{aligned} &\frac{s(s-1)}{r^2} + 4\nu^2 r^2 + 16\nu^2 r^2\frac{\Lambda''}{\Lambda} + 4\nu\left(2s+1-4\nu r^2\right)\frac{\Lambda'}{\Lambda} - 2\nu(1+2s) + \frac{2}{r}\left(\frac{s}{r} + 4\nu r\frac{\Lambda'}{\Lambda} - 2\nu r\right) + \\ &+\frac{2m\mathrm{E}}{\hbar^2} + \frac{1+\kappa-\tau^2}{r^2} - \frac{r^2}{4\sigma_r^4} = 0, \\ &\frac{s(s+1)+1+\kappa-\tau^2}{r^2} + \left(16\sigma_r^4\nu^2 - 1\right)\frac{r^2}{4\sigma_r^4} + \\ &+8\nu\left[2\nu r^2\frac{\Lambda''}{\Lambda} + \left(s+\frac{3}{2}-2\nu r^2\right)\frac{\Lambda'}{\Lambda}\right] + \frac{2m\mathrm{E}}{\hbar^2} - 2\nu(3+2s) = 0. \end{aligned} \quad \text{(A.13)}$$

Equation (A.13) can be satisfied by setting the free parameters $\mathrm{E}, \nu, \tau$, and imposing a condition on the unknown function $\Lambda$

$$2\nu r^2\frac{\Lambda''}{\Lambda} + \left(s+\frac{3}{2}-2\nu r^2\right)\frac{\Lambda'}{\Lambda} = -n = const, \quad \Lambda\left(2\nu r^2\right) = L_n^{(s+1/2)}\left(2\nu r^2\right), \quad \text{(A.14)}$$

where $L_n^{(s+1/2)}$ are generalized Laguerre polynomials. Substituting (A.14) into (A.13), we obtain

$$\frac{s(s+1)+1+\kappa-\tau^2}{r^2} + \left(16\sigma_r^4\nu^2 - 1\right)\frac{r^2}{4\sigma_r^4} + \frac{2m\mathrm{E}}{\hbar^2} - 2\nu(3+4n+2s) = 0,$$

hence

$$\nu = \frac{1}{4\sigma_r^2}, \qquad s = \kappa, \qquad \tau^2 = (\kappa+1)^2, \qquad \mathrm{E} = \frac{\hbar^2}{2m\sigma_r^2}\left(2n+\kappa+\frac{3}{2}\right). \quad \text{(A.15)}$$

Thus, the wave function (2.2), taking into account (A.11), (A.14), and (A.15), takes the form

$$\Psi(r,\theta,\phi)=\frac{N}{\sqrt{\sin\theta}}r^{\kappa}L_n^{(\kappa+1/2)}\left(2\nu r^2\right)e^{-\nu r^2+i\left(\ell\phi+\tau\theta-\frac{\mathrm{E}}{\hbar}t\right)}, \tag{A.16}$$

where $N$ is the normalization factor, determined as

$$1=\int_{\mathbb{R}^3}|\Psi|^2 d^3r=N^2\int_0^{2\pi}d\phi\int_0^{\pi}d\theta\int_0^{+\infty}r^{2\kappa+2}\left[L_n^{(\kappa+1/2)}\left(2\nu r^2\right)\right]^2 e^{-2\nu r^2}dr=$$
$$=\frac{\pi^2N^2}{(2\nu)^{\kappa+3/2}}\int_0^{+\infty}\bar{r}^{\kappa+1/2}e^{-\mu}\left[L_n^{(\kappa+1/2)}(\bar{r})\right]^2 d\bar{r}=\frac{\pi^2N^2}{(2\nu)^{\kappa+3/2}n!}\Gamma\left(n+\kappa+\frac{3}{2}\right),$$
$$N^2=\frac{2^{1+2\kappa+2n}(2\nu)^{\kappa+3/2}n!(n+\kappa)!}{\pi^2\sqrt{\pi}(2n+2\kappa+1)!}=\frac{2^{\kappa+2n}n!(n+\kappa)!}{\sigma_r^{2\kappa+3}\pi^2\sqrt{2\pi}(2n+2\kappa+1)!},$$
$$N=\frac{2^n}{\pi\sigma_r^{\kappa+1}}\sqrt{\frac{2^{\kappa}n!(n+\kappa)!}{\sigma_r\sqrt{2\pi}(2n+2\kappa+1)!}}, \tag{A.17}$$

where we have taken into account

$$\Gamma\left(n+\kappa+\frac{3}{2}\right)=\Gamma\left(n+\kappa+1+\frac{1}{2}\right)=\frac{\Gamma(2n+2\kappa+2)}{\Gamma(n+\kappa+1)}\frac{\sqrt{\pi}}{2^{1+2n+2\kappa}}=\frac{(2n+2\kappa+1)!}{(n+\kappa)!}\frac{\sqrt{\pi}}{2^{1+2n+2\kappa}}.$$

The expression for the quantum potential Q is obtained from the Hamilton-Jacobi equation

$$\mathrm{Q}=-\hbar\frac{\partial\varphi}{\partial t}-\frac{m}{2}\left|\langle\vec{v}\rangle\right|^2-U=\mathrm{E}-\frac{\hbar^2}{2m}\left(\frac{\ell^2}{r^2\sin^2\theta}+\frac{\tau^2}{r^2}\right)-\frac{\hbar^2}{8m}\left(\frac{r^2}{\sigma_r^4}-\frac{4\kappa+3}{r^2}+\frac{1-4\ell^2}{r^2\sin^2\theta}\right)=$$
$$=\frac{\hbar^2}{2m\sigma_r^2}\left(2n+\kappa+\frac{3}{2}\right)-\frac{\hbar^2}{8m}\left(\frac{4\ell^2}{r^2\sin^2\theta}+\frac{4\kappa^2+8\kappa+4}{r^2}+\frac{r^2}{\sigma_r^4}-\frac{4\kappa+3}{r^2}+\frac{1-4\ell^2}{r^2\sin^2\theta}\right),$$
$$\mathrm{Q}=-\frac{\hbar^2}{8m\sigma_r^2}\left\{\frac{r^2}{\sigma_r^2}+\left[4\kappa(\kappa+1)+1\right]\frac{\sigma_r^2}{r^2}+\frac{\sigma_r^2}{r^2\sin^2\theta}-8n-4\kappa-6\right\}. \tag{A.18}$$

Theorem 1 is proven.

***Proof of Theorem 2***

The proof follows the same approach as the proof of Theorem 1. We perform the following calculations

$$\left(\hat{\mathrm{p}}-q\vec{A}\right)^2\Psi=\left(\hat{\mathrm{p}}-q\vec{A}\right)\left(\hat{\mathrm{p}}\Psi-q\vec{A}\Psi\right)=\hat{\mathrm{p}}^2\Psi-q\Psi\hat{\mathrm{p}}\vec{A}-2q\vec{A}\hat{\mathrm{p}}\Psi+q^2\left|\vec{A}\right|^2\Psi=$$
$$=\hat{\mathrm{p}}^2\Psi-2q\vec{A}\hat{\mathrm{p}}\Psi+q^2\left|\vec{A}\right|^2\Psi, \tag{A.19}$$

where we have taken into account that $\mathrm{div}_r\vec{A}=0$. Considering the form of the vector potential $\vec{A}$, expression (A.19) becomes

$$\frac{\left(\hat{\mathrm{p}}-q\vec{A}\right)^2\Psi}{\Psi}=-\hbar^2\frac{\Delta_r\Psi}{\Psi}-2i\hbar^2\frac{\ell}{r\sin\theta}\vec{e}_\phi\frac{\nabla_r\Psi}{\Psi}+\frac{\hbar^2\ell^2}{\rho^2}. \tag{A.20}$$

According to the $\Psi$ - model, we seek the solution in the form

$$\Psi(r,\theta,\phi,t)=\frac{R(r)}{\sqrt{\sin\theta}}\Upsilon\left(\phi+\frac{\ell}{\tau}\operatorname{ctg}\theta\right)e^{i\left(\tau\theta-\frac{\mathrm{E}}{\hbar}t\right)}, \tag{A.21}$$

which gives

$$\vec{e}_\phi\nabla_r\Psi=\frac{Re^{i\left(\tau\theta-\frac{\mathrm{E}}{\hbar}t\right)}\Upsilon'}{r\sin\theta\sqrt{\sin\theta}},\qquad \vec{e}_\phi\frac{\nabla_r\Psi}{\Psi}=\frac{1}{r\sin\theta}\frac{\Upsilon'}{\Upsilon}, \tag{A.22}$$

$$\begin{aligned}\Delta_r\Psi=\Upsilon\frac{e^{i\left(\tau\theta-\frac{\mathrm{E}}{\hbar}t\right)}}{r\sqrt{\sin\theta}}\frac{\partial^2}{\partial r^2}(rR)+e^{-i\frac{\mathrm{E}}{\hbar}t}\frac{R}{r^2\sin\theta}\frac{\partial}{\partial\theta}\left(\sin\theta\frac{\partial}{\partial\theta}\frac{e^{i\tau\theta}}{\sqrt{\sin\theta}}\Upsilon\right)+\\ +e^{i\left(\tau\theta-\frac{\mathrm{E}}{\hbar}t\right)}\frac{R}{r^2\sin^2\theta}\frac{1}{\sqrt{\sin\theta}}\frac{\partial^2\Upsilon}{\partial\phi^2}.\end{aligned} \tag{A.23}$$

Using the results (A.2) and (A.4), expression (A.23) becomes

$$\begin{aligned}\frac{\Delta_r\Psi}{\Psi}=\frac{R''}{R}+\frac{2}{r}\frac{R'}{R}+\frac{1}{r^2\sin^2\theta}\left(\frac{\ell^2}{\tau^2\sin^2\theta}+1\right)\frac{\Upsilon''}{\Upsilon}+\frac{2\ell}{r^2\sin^2\theta}\left(\frac{\cos\theta}{\tau\sin\theta}-i\right)\frac{\Upsilon'}{\Upsilon}+\\ +\frac{1}{4r^2\sin^2\theta}+\frac{1-4\tau^2}{4r^2}.\end{aligned} \tag{A.24}$$

Substituting (A.22) and (A.24) into (A.20), we obtain

$$\begin{aligned}-\frac{\left(\hat{\mathrm{p}}-q\vec{A}\right)^2\Psi}{\hbar^2\Psi}=\frac{R''}{R}+\frac{2}{r}\frac{R'}{R}+\frac{1}{r^2\sin^2\theta}\left(\frac{\ell^2}{\tau^2\sin^2\theta}+1\right)\frac{\Upsilon''}{\Upsilon}+\frac{2\ell\cos\theta}{\tau r^2\sin^3\theta}\frac{\Upsilon'}{\Upsilon}+\\ +\frac{1-4\ell^2}{4r^2\sin^2\theta}+\frac{1-4\tau^2}{4r^2}.\end{aligned} \tag{A.25}$$

From the electromagnetic Schrödinger equation for function (A.21), it follows that the potential $U$ can be expressed as

$$\begin{aligned}U=\mathrm{E}-\frac{1}{2m}\frac{\left(\hat{\mathrm{p}}-q\vec{A}\right)^2\Psi}{\Psi}=\frac{\hbar^2}{8m\sigma_r^2}\left[\frac{r^2}{\sigma_r^2}-(4\kappa+3)\frac{\sigma_r^2}{r^2}+\left(1-4\ell^2\right)\frac{\sigma_r^2}{r^2\sin^2\theta}\right]=\\ =\mathrm{E}+\frac{\hbar^2}{2m}\left[\frac{R''}{R}+\frac{2}{r}\frac{R'}{R}+\frac{1}{r^2\sin^2\theta}\left(\frac{\ell^2}{\tau^2\sin^2\theta}+1\right)\frac{\Upsilon''}{\Upsilon}+\frac{2\ell\cos\theta}{\tau r^2\sin^3\theta}\frac{\Upsilon'}{\Upsilon}+\frac{1-4\ell^2}{4r^2\sin^2\theta}+\frac{1-4\tau^2}{4r^2}\right].\end{aligned} \tag{A.26}$$

Expression (A.26) coincides with (A.7), therefore

$$\Psi(r,\theta,t)=\frac{N}{\sqrt{\sin\theta}}r^\kappa L_n^{(\kappa+1/2)}\left(2\nu r^2\right)e^{-\nu r^2+i\left(\pm\tau\theta-\frac{\mathrm{E}}{\hbar}t\right)}, \tag{A.27}$$

where $N$ is the normalization factor (A.17). Theorem 2 is proven.

***Proof of Theorem 3***

Substitute solution (3.3) into equation (2.3)

$$\frac{2mU}{\hbar^2} = \frac{2m\mathrm{E}}{\hbar^2} + \frac{\Delta\Psi}{\Psi} = \frac{2m}{\hbar^2}\frac{\hbar^2}{8m\sigma_r^2}\left(\frac{r^2}{\sigma_r^2} - \frac{4\varepsilon^2\sigma_r^2}{r^2\sin^2\theta}\right) = \\ = \frac{2m\mathrm{E}}{\hbar^2} + \frac{R''}{R} + \frac{2}{r}\frac{R'}{R} + \frac{1}{r^2}\left(\operatorname{ctg}\theta\frac{\Theta'}{\Theta} + \frac{\Theta''}{\Theta} - \frac{\ell^2}{\sin^2\theta}\right). \tag{A.28}$$

Transform the angular part in equation (A.28) by introducing the variable substitution

$$\Theta(\theta) = P(\cos\theta) \Rightarrow \Theta' = -P'\sin\theta,\ \Theta'' = -P'\cos\theta + P''\sin^2\theta, \tag{A.29}$$

we obtain,

$$\frac{2m\mathrm{E}}{\hbar^2} + \frac{R''}{R} + \frac{2}{r}\frac{R'}{R} - \frac{r^2}{4\sigma_r^4} + \frac{1}{r^2}\left(\sin^2\theta\frac{P''}{P} - 2\cos\theta\frac{P'}{P} + \frac{\varepsilon^2 - \ell^2}{\sin^2\theta}\right) = 0,$$

$$\sin^2\theta\frac{P''}{P} - 2\cos\theta\frac{P'}{P} - \frac{\ell^2 - \varepsilon^2}{\sin^2\theta} = -r^2\left(\frac{R''}{R} + \frac{2}{r}\frac{R'}{R} + \frac{2m\mathrm{E}}{\hbar^2} - \frac{r^2}{4\sigma_r^4}\right) = -\lambda = const, \tag{A.30}$$

hence

$$\begin{cases} R'' + \dfrac{2}{r}R' + \left(\dfrac{2m\mathrm{E}}{\hbar^2} - \dfrac{\lambda}{r^2} - \dfrac{r^2}{4\sigma_r^4}\right)R = 0, \\ \sin^2\theta P'' - 2\cos\theta P' + \left(\lambda - \dfrac{\ell^2 - \varepsilon^2}{\sin^2\theta}\right)P = 0. \end{cases} \tag{A.31}$$

The solution to the first equation in system (A.31) has the form (A.10)-(A.11), i.e.

$$R(r) = r^s L_n^{(s+1/2)}\left(2\nu r^2\right)e^{-\nu r^2},\ \lambda = s(s+1). \tag{A.32}$$

The solution to the second equation (A.31) is expressed in terms of associated Legendre polynomials

$$P_s^l(\cos\theta), \quad l^2 = \ell^2 - \varepsilon^2, \quad 0 \le l \le s. \tag{A.33}$$

Taking into account (A.32) and (A.33), the solution to the Schrödinger equation takes the form

$$\Psi(\vec{r},t) = Mr^s L_n^{(s+1/2)}\left(2\nu r^2\right)P_s^l(\cos\theta)e^{-\nu r^2 + i\left(\ell\phi - \frac{\mathrm{E}}{\hbar}t\right)}. \tag{A.34}$$

Find the normalization coefficient $M$ .

$$1=\int_{\mathbb{R}^3}|\Psi|^2 d^3r = M^2\int_0^{2\pi}d\phi\int_0^{\pi}\sin\theta\left[P_s^l(\cos\theta)\right]^2 d\theta\int_0^{+\infty} r^{2s+2}\left[L_n^{(s+1/2)}\left(2\nu r^2\right)\right]^2 e^{-2\nu r^2}dr =$$

$$=\frac{M^2}{2\pi^2\left[N_n^{(s)}\right]^2}\int_0^{2\pi}d\phi\int_{-1}^{1}\left[P_s^l(x)\right]^2 dx = \frac{M^2}{\pi\left[N_n^{(s)}\right]^2}\frac{2(s+l)!}{(2s+1)(s-l)!},$$

$$M^2=\pi\left[N_n^{(s)}\right]^2\frac{(2s+1)(s-l)!}{2(s+l)!}=\frac{2^{s+2n}n!(n+s)!(2s+1)(s-l)!}{\sigma_r^{2s+3}2\pi\sqrt{2\pi}(2n+2s+1)!(s+l)!}, \tag{A.35}$$

where expression (A.17) and the relation are taken into account

$$\int_{-1}^{1}P_s^l(x)P_{s'}^l(x)dx=\frac{2(s'+l)!}{(2s'+1)(s'-l)!}\delta_{s,s'}. \tag{A.36}$$

Calculate the quantum potential. By analogy with expression (A.18), we obtain

$$\mathrm{Q}=-\hbar\frac{\partial\varphi}{\partial t}-\frac{m}{2}\left|\langle\vec{v}\rangle\right|^2-U=\mathrm{E}-\frac{\hbar^2\ell^2}{2mr^2\sin^2\theta}-\frac{\hbar^2}{8m\sigma_r^2}\left(\frac{r^2}{\sigma_r^2}-\frac{4\varepsilon^2\sigma_r^2}{r^2\sin^2\theta}\right)=$$

$$=\frac{\hbar^2}{2m\sigma_r^2}\left(2n+s+\frac{3}{2}\right)-\frac{\hbar^2}{8m\sigma_r^2}\left[4\left(\ell^2-\varepsilon^2\right)\frac{\sigma_r^2}{r^2\sin^2\theta}+\frac{r^2}{\sigma_r^2}\right]= \tag{A.37}$$

$$=-\frac{\hbar^2}{8m\sigma_r^2}\left(\frac{4l^2\sigma_r^2}{r^2\sin^2\theta}+\frac{r^2}{\sigma_r^2}-8n-4s-6\right).$$

Theorem 3 is proved.

***Proof of Lemma 1***

Perform direct calculation of the integrals. For the first expression, we obtain

$$\int_{S^2}Y_s^{(\ell,l)}Y_{s'}^{(\ell',l)*}ds=\int_0^{2\pi}d\phi\int_0^{\pi}Y_s^{(\ell,l)}Y_{s'}^{(\ell',l)}\sin\theta d\theta=\sqrt{\frac{(2s+1)(s-l)!}{4\pi(s+l)!}}\sqrt{\frac{(2s'+1)(s'-l)!}{4\pi(s'+l)!}}\times$$

$$\times\int_0^{2\pi}e^{i(\ell-\ell')\phi}d\phi\int_{-1}^{1}P_s^l(x)P_{s'}^l(x)dx=\delta_{\ell,\ell'}\delta_{s,s'}, \tag{A.38}$$

where (A.36) is taken into account. Check the second relation

$$\int_{S^2}\frac{1}{\sin^2\theta}Y_s^{(\ell,l)}Y_s^{(\ell',l')*}ds=\sqrt{\frac{(2s+1)(s-l)!}{4\pi(s+l)!}}\sqrt{\frac{(2s+1)(s-l')!}{4\pi(s+l')!}}\int_0^{2\pi}e^{i(\ell-\ell')\phi}d\phi\int_{-1}^{1}\frac{P_s^l(x)P_s^{l'}(x)}{1-x^2}dx=$$

$$=(2s+1)\sqrt{\frac{(s-l)!(s+l')!}{(s+l)!(s-l')!}}\frac{\delta_{\ell,\ell'}\delta_{l,l'}}{2l'}=\frac{2s+1}{2l}\delta_{\ell,\ell'}\delta_{l,l'}, \tag{A.39}$$

where it is taken into account

$$\int_{-1}^{1}\frac{P_s^l(x)P_s^{l'}(x)}{1-x^2}dx=\frac{(s+l')!}{l'(s-l')!}\delta_{l,l'}. \tag{A.40}$$

Lemma 1 is proved.

***Proof of Theorem 4***

By analogy with the proof of Theorem 3, we will look for a solution in the form

$$\Psi(r,\theta,t) = R(r)\Theta(\theta)e^{i\left(\kappa\phi - \frac{\mathrm{E}}{\hbar}t\right)}. \tag{A.40}$$

Taking into account (A.40), expression (A.26) takes the form

$$-\frac{\left(\hat{\mathrm{p}} - q\vec{A}\right)^2 \Psi}{\hbar^2 \Psi} = \frac{R''}{R} + \frac{2}{r}\frac{R'}{R} + \frac{1}{r^2}\left(\frac{\Theta''}{\Theta} + \frac{\Theta'}{\Theta}\operatorname{ctg}\theta\right) - \frac{(\kappa+\ell)^2}{r^2\sin^2\theta},$$

hence

$$\frac{2mU}{\hbar^2} = \frac{2m\mathrm{E}}{\hbar^2} - \frac{\left(\hat{\mathrm{p}} - q\vec{A}\right)^2 \Psi}{\Psi\hbar^2} = \frac{r^2}{4\sigma_r^4} - \frac{\varepsilon^2}{r^2\sin^2\theta} = \tag{A.41}$$

$$= \frac{2m\mathrm{E}}{\hbar^2} + \frac{R''}{R} + \frac{2}{r}\frac{R'}{R} + \frac{1}{r^2}\left[\frac{\Theta''}{\Theta} + \operatorname{ctg}\theta\frac{\Theta'}{\Theta} - \frac{(\ell+\kappa)^2}{\sin^2\theta}\right].$$

Expression (A.41) coincides with (A.28), therefore,

$$\Psi(\vec{r},t) = Mr^s L_n^{(s+1/2)}\left(2\nu r^2\right)P_s^l(\cos\theta)e^{-\nu r^2 + i\kappa\phi - i\frac{\mathrm{E}}{\hbar}t},\ \ l^2 = (\ell+\kappa)^2 - \varepsilon^2,\ 0 \le l \le s, \tag{A.42}$$

where $M$ is the normalization factor (A.35). Theorem 4 is proved.

**Appendix B**

To normalize the function (2.11), it is necessary

$$\left|\Psi^{(\ell)}\right|^2 = \Psi^{(\ell)}\Psi^{(\ell)*} = \left(c_1 e^{i\ell\phi} + c_2 e^{-i\ell\phi}\right)\Psi_n^{(\kappa,0,\tau)}\left(c_1^* e^{-i\ell\phi} + c_2^* e^{i\ell\phi}\right)\Psi_n^{(\kappa,0,\tau)*} =$$

$$= \left|\Psi_n^{(\kappa,0,\tau)}\right|^2\left(|c_1|^2 + |c_2|^2 + c_1 c_2^* e^{2i\ell\phi} + c_2 c_1^* e^{-2i\ell\phi}\right),$$

$$1 = \int_{\mathbb{R}^3}\left|\Psi^{(\ell)}\right|^2 d^3r = \int_0^{2\pi}\left[|c_1|^2 + |c_2|^2 + c_1 c_2^* e^{2i\ell\phi} + c_2 c_1^* e^{-2i\ell\phi}\right]d\phi\int_0^{\pi}\sin\theta d\theta\int_0^{+\infty} r^2\left|\Psi_n^{(\kappa,0,\tau)}\right|^2 dr =$$

$$= \left(|c_1|^2 + |c_2|^2\right)\int_0^{2\pi}d\phi\int_0^{\pi}\sin\theta d\theta\int_0^{+\infty} r^2\left|\Psi_n^{(\kappa,0,\tau)}\right|^2 dr + c_1 c_2^*\int_0^{2\pi}e^{2i\ell\phi}d\phi\int_0^{\pi}\sin\theta d\theta\int_0^{+\infty} r^2\left|\Psi_n^{(\kappa,0,\tau)}\right|^2 dr +$$

$$+ c_2 c_1^*\int_0^{2\pi}e^{-2i\ell\phi}d\phi\int_0^{\pi}\sin\theta d\theta\int_0^{+\infty} r^2\left|\Psi_n^{(\kappa,0,\tau)}\right|^2 dr,$$

$$|c_1|^2 + |c_2|^2 + c_1 c_2^*\frac{e^{4\pi i\ell} - 1}{4\pi i\ell} - c_2 c_1^*\frac{e^{-4\pi i\ell} - 1}{4\pi i\ell} = 1. \tag{B.1}$$

If $\ell \in \mathbb{Z}$, then $e^{\pm 4i\pi\ell} = 1$ and the normalization condition takes the form $|c_1|^2 + |c_2|^2 = 1$.

Calculate the intrinsic magnetic moment for distribution (2.10) with field (1.17), we obtain

$$\vec{\mu}_S^{(1)} = \frac{1}{2}\int_{\mathbb{R}^3}\left(\vec{r}\times\vec{J}\right)d^3r = \mu_B\int_0^{2\pi}d\phi\int_0^{\pi}\sin\theta d\theta\int_0^{+\infty}f_1r^2\left[\tau\left(\vec{e}_r\times\vec{e}_\theta\right)+\frac{\ell}{\sin\theta}\left(\vec{e}_r\times\vec{e}_\phi\right)\right]dr =$$
$$= \mu_B\int_0^{2\pi}d\phi\int_0^{\pi}\sin\theta d\theta\int_0^{+\infty}r^2f_1\left(\vec{r}\right)\left(-\frac{\ell}{\sin\theta}\vec{e}_\theta+\tau\vec{e}_\phi\right)dr = \tag{B.2}$$
$$= -\mu_B\ell\int_0^{2\pi}d\phi\int_0^{\pi}d\theta\int_0^{+\infty}\vec{e}_\theta r^2f_1\left(\vec{r}\right)dr+\mu_B\tau\int_0^{2\pi}d\phi\int_0^{\pi}\sin\theta d\theta\int_0^{+\infty}\vec{e}_\phi r^2f_1\left(\vec{r}\right)dr = \vec{I}_1+\vec{I}_2.$$

Find each integral $\vec{I}_1,\vec{I}_2$ separately

$$\vec{I}_1 = -\mu_B\ell\int_0^{2\pi}d\phi\int_0^{\pi}\begin{pmatrix}\cos\theta\cos\phi\\ \cos\theta\sin\phi\\ -\sin\theta\end{pmatrix}d\theta\int_0^{+\infty}r^2f_1\left(\vec{r}\right)dr =$$
$$= -\mu_B\ell\left[N_n^\kappa\right]^2\begin{pmatrix}\int_0^{2\pi}\cos\phi d\phi\int_0^{\pi}\operatorname{ctg}\theta d\theta\\ \int_0^{2\pi}\sin\phi d\phi\int_0^{\pi}\operatorname{ctg}\theta d\theta\\ -\int_0^{2\pi}d\phi\int_0^{\pi}d\theta\end{pmatrix}\int_0^{+\infty}r^{2\kappa+2}\left[L_n^{(\kappa+1/2)}\left(2\nu r^2\right)\right]^2e^{-2\nu r^2}dr = \frac{\mu_B\ell}{2\pi^2}\begin{pmatrix}0\\0\\2\pi^2\end{pmatrix}, \tag{B.3}$$

$$\vec{I}_1 = \mu_B\vec{e}_z, \tag{B.4}$$

where (A.17) has been accounted for,

$$\vec{I}_2 = \mu_B\tau\left[N_n^\kappa\right]^2\int_0^{2\pi}\begin{pmatrix}-\sin\phi\\ \cos\phi\\ 0\end{pmatrix}d\phi\int_0^{\pi}d\theta\int_0^{+\infty}r^{2\kappa+2}\left[L_n^{(\kappa+1/2)}\left(2\nu r^2\right)\right]^2e^{-2\nu r^2}dr = \vec{\theta}, \tag{B.5}$$

Substituting (B.4) and (B.5) into (B.2), we obtain $\vec{\mu}_S^{(1)} = \mu_B\ell\vec{e}_z$.

For superposition (2.12) $\ell = 0$ (see Fig. 4 on the right). By analogy with (B.3), we get $\vec{I}_1 = \vec{\theta}$. Substitution of $f_{n,\Sigma+_\ell}^{(\ell)} \sim \cos^2\left(\ell\phi\right)$ into the integral $\vec{I}_2$ gives the result

$$\vec{I}_2 = 2\mu_B\tau\left[N_n^\kappa\right]^2\int_0^{2\pi}\begin{pmatrix}-\sin\phi\cos^2\left(\ell\phi\right)\\ \cos\phi\cos^2\left(\ell\phi\right)\\ 0\end{pmatrix}d\phi\int_0^{\pi}\frac{d\theta}{\sin\theta}\int_0^{+\infty}r^{2\kappa+2}\left[L_n^{(\kappa+1/2)}\left(2\nu r^2\right)\right]^2e^{-2\nu r^2}dr = \vec{\theta}, \tag{B.6}$$

where it is taken into account that $\int_0^{2\pi}(...)d\phi = 0$. When substituting $f_{n,\Sigma-_\ell}^{(\ell)} \sim \sin^2\left(\ell\phi\right)$, a result similar to (B.6) is obtained. Thus, $\vec{\mu}_S^{(2)} = \vec{\theta}$.

The magnetic moment for superposition (2.19) is determined by the flow at $\tau=0$, i.e., $\dfrac{\hbar\ell\vec{e}_\phi}{mr\sin\theta}$. Therefore, the integral $\vec{I}_1$ for the distribution function $f^{(\tau)}_{n,\Sigma+_\tau}$ (2.20) takes the form

$$\vec{I}_1=-2\mu_{\rm B}\ell\left[N_n^{\kappa}\right]^2\begin{pmatrix}\int\limits_0^{2\pi}\cos\phi d\phi\int_0^{\pi}\cos^2(\tau\theta)\,\mathrm{ctg}\,\theta d\theta\\ \int\limits_0^{2\pi}\sin\phi d\phi\int_0^{\pi}\cos^2(\tau\theta)\,\mathrm{ctg}\,\theta d\theta\\ -\int\limits_0^{2\pi}d\phi\int_0^{\pi}\cos^2(\tau\theta)d\theta\end{pmatrix}\int\limits_0^{+\infty}r^{2\kappa+2}\left[L_n^{(\kappa+1/2)}\left(2\nu r^2\right)\right]^2e^{-2\nu r^2}dr=2\frac{\mu_{\rm B}\ell}{2\pi^2}\begin{pmatrix}0\\0\\\pi^2\end{pmatrix},$$

$$\vec{I}_1=\mu_{\rm B}\ell\vec{e}_z. \tag{B.7}$$

The result (B.7) is also valid for the function $f^{(\tau)}_{n,\Sigma-_\tau}$. Due to the condition $\tau=0$ the integral $\vec{I}_2=\vec{\theta}$. Finally, $\vec{\mu}_{\rm S}^{(3)}=\mu_{\rm B}\ell\vec{e}_z$.

Calculate $\vec{\mu}_s$ for system (3.4). Using expression (B.2), since $\tau=0$, it is sufficient to calculate only $\vec{I}_1$, we obtain

$$\vec{I}_1=-\mu_{\rm B}\ell\int\limits_0^{2\pi}d\phi\int\limits_0^{\pi}\begin{pmatrix}\cos\theta\cos\phi\\ \cos\theta\sin\phi\\ -\sin\theta\end{pmatrix}d\theta\int\limits_0^{+\infty}r^2f_1(\vec{r})dr=$$

$$=-\mu_{\rm B}\ell\left[N_{n,s}^{(\pm l)}\right]^2\begin{pmatrix}\int\limits_0^{2\pi}\cos\phi d\phi\int_0^{\pi}\cos\theta\left[P_s^{\pm l}(\cos\theta)\right]^2d\theta\\ \int\limits_0^{2\pi}\sin\phi d\phi\int_0^{\pi}\cos\theta\left[P_s^{\pm l}(\cos\theta)\right]^2d\theta\\ -\int\limits_0^{2\pi}d\phi\int_0^{\pi}\sin\theta\left[P_s^{\pm l}(\cos\theta)\right]^2d\theta\end{pmatrix}\int\limits_0^{+\infty}r^{2s+2}\left[L_n^{(s+1/2)}\left(2\nu r^2\right)\right]^2e^{-2\nu r^2}dr=$$

$$=\vec{e}_z\frac{\mu_{\rm B}\ell\left[N_{n,s}^{(\pm l)}\right]^2}{\pi\left[N_n^{(s)}\right]^2}\int_{-1}^{1}\left[P_s^{\pm l}(x)\right]^2dx=\mu_{\rm B}\ell\vec{e}_z, \tag{B.8}$$

where expressions (A.17), (A.35) and (A.36) are taken into account.

## Acknowledgments

The study was conducted under the state assignment of Lomonosov Moscow State University.

## Declarations

**Funding** Not applicable

**Availability of data and material** This manuscript has no associated data.

**Competing interests** The author declare that he has no conflict of interest.

**Authors Contributions** E.E. Perepelkin: Conceptualization, Investigation, Visualization, Writing – original draft. B.I. Sadovnikov: Conceptualization, Supervision, Validation. N.G. Inozemtseva: Formal analysis, Validation, Writing – review & editing. A.S. Medvedev: Investigation, Software, Visualization, Writing – original draft.